\pdfoutput=1
\documentclass[11pt,twoside,a4paper,cmspaper,final,collab]{cms-tdr}
\def\svnVersion{9470f9a}\def\svnDate{2023/11/08}\def\cmsCernNoTag{CERN-EP-2023-090}\def\cmsCernDate{\today}\def\cmsMessage{Published in Physics Letters B as \href{http://dx.doi.org/10.1016/j.physletb.2023.138290}{\doi{10.1016/j.physletb.2023.138290}.}}
\begin{document}\cmsNoteHeader{TOP-22-013}

\ifthenelse{\boolean{cms@external}}{%
    \providecommand{\cmsUpperLeft}{upper left\xspace}
    \providecommand{\cmsUpperRight}{upper right\xspace}
    \providecommand{\cmsLower}{lower\xspace}
    \providecommand{\cmsSpace}[1]{\hfill}
}{%
    \providecommand{\cmsUpperLeft}{left\xspace}
    \providecommand{\cmsUpperRight}{middle\xspace}
    \providecommand{\cmsLower}{right\xspace}
    \providecommand{\cmsSpace}[1]{\hspace{#1}}
}

\providecommand{\cmsTable}[1]{\resizebox{\textwidth}{!}{#1}}
\newlength\cmsTabSkip\setlength{\cmsTabSkip}{2ex}

\newcommand{\sqrts}{\ensuremath{\sqrt{s}=13\TeV}\xspace}
\newcommand{\pp}{\ensuremath{\Pp\Pp}\xspace}

\newcommand{\pmasym}[2]{\ensuremath{\,^{#1}_{#2}}}
\newcommand{\fb}{\unit{fb}}
\newcommand{\sigobs}{\ensuremath{5.6}\xspace}
\newcommand{\sigexp}{\ensuremath{4.9}\xspace}
\newcommand{\xsecpred}{\ensuremath{13.4\pmasym{+1.0}{-1.8}\fb}\xspace}
\newcommand{\xsecmeas}{\ensuremath{17.7\pmasym{+3.7}{-3.5}\stat\pmasym{+2.3}{-1.9}\syst\fb}\xspace}
\newcommand{\ttwpred}{\ensuremath{722\pm74\fb}\xspace}
\newcommand{\ttzpred}{\ensuremath{859\pm80\fb}\xspace}

\newcommand{\sigeta}{\ensuremath{\eta}\xspace}
\newcommand{\abseta}{\ensuremath{\abs{\sigeta}}\xspace}
\newcommand{\Deta}{\ensuremath{\Delta\sigeta}\xspace}
\newcommand{\Dphi}{\ensuremath{\Delta\phi}\xspace}
\newcommand{\Irel}{\ensuremath{I_{\text{rel}}}\xspace}
\newcommand{\PQj}{{\HepParticle{j}{}{}}\xspace}
\newcommand{\Njets}{\ensuremath{N_{\PQj}}\xspace}
\newcommand{\Nbjets}{\ensuremath{N_{\PQb}}\xspace}
\newcommand{\mll}{\ensuremath{m(\Pell\Pell)}\xspace}
\newcommand{\mthreel}{\ensuremath{m(\Pell\Pell\Pell)}\xspace}
\newcommand{\mZ}{\ensuremath{m_{\PZ}}\xspace}
\newcommand{\logsoverb}{\ensuremath{\log_{10}(\mathrm{S}/\mathrm{B})}\xspace}

\newcommand{\tttt}{\ensuremath{\ttbar\ttbar}\xspace}
\newcommand{\sigtttt}{\ensuremath{\sigma(\tttt)}\xspace}
\newcommand{\ttbb}{\ensuremath{\ttbar\bbbar}\xspace}
\newcommand{\ttt}{\ensuremath{\PQt\PQt\PQt}\xspace}
\newcommand{\ttZ}{\ensuremath{\ttbar\PZ}\xspace}
\newcommand{\ttW}{\ensuremath{\ttbar\PW}\xspace}
\newcommand{\ttH}{\ensuremath{\ttbar\PH}\xspace}
\newcommand{\ttX}{\ensuremath{\ttbar\PX}\xspace}
\newcommand{\WW}{\ensuremath{\PW\PW}\xspace}
\newcommand{\WZ}{\ensuremath{\PW\PZ}\xspace}
\newcommand{\ZZ}{\ensuremath{\PZ\PZ}\xspace}
\newcommand{\Zjets}{\ensuremath{\PZ\text{+jets}}\xspace}
\newcommand{\tZq}{\ensuremath{\PQt\PZ\PQq}\xspace}
\newcommand{\otherT}{\ensuremath{\text{other~}\PQt}\xspace}
\newcommand{\VVV}{\ensuremath{\PV\PV(\PV)}\xspace}
\newcommand{\Xgamma}{\ensuremath{\PX\PGg}\xspace}

\newcommand{\DeepJet}{\ensuremath{\textsc{DeepJet}}\xspace}
\newcommand{\TMVA}{\ensuremath{\textsc{tmva}}\xspace}
\newcommand{\DpJ}{\ensuremath{\mathrm{DJ}}\xspace}
\newcommand{\mtop}[1]{\ensuremath{m(\PQt_#1)}\xspace}
\newcommand{\mW}[1]{\ensuremath{m(\PW_#1)}\xspace}

\newcommand{\twol}{\ensuremath{2\Pell}\xspace}
\newcommand{\threel}{\ensuremath{3\Pell}\xspace}
\newcommand{\fourl}{\ensuremath{4\Pell}\xspace}

\newcommand{\SRtwo}{\ensuremath{\text{SR-}\twol}\xspace}
\newcommand{\SRthree}{\ensuremath{\text{SR-}\threel}\xspace}
\newcommand{\SRfour}{\ensuremath{\text{SR-}\fourl}\xspace}
\newcommand{\CRtwoHigh}{\ensuremath{\text{CR-}\twol\text{-45j2\PQb}}\xspace}
\newcommand{\CRtwoLow}{\ensuremath{\text{CR-}\twol\text{-23j1\PQb}}\xspace}
\newcommand{\CRthreeLow}{\ensuremath{\text{CR-}\threel\text{-2j1\PQb}}\xspace}
\newcommand{\CRthreeZ}{\ensuremath{\text{CR-}\threel\text{-\PZ}}\xspace}
\newcommand{\CRfourZ}{\ensuremath{\text{CR-}\fourl\text{-\PZ}}\xspace}
\newcommand{\mumu}{\ensuremath{\PGm\PGm}\xspace}

\newcommand{\likeli}{\ensuremath{L}\xspace}
\newcommand{\sigstr}{\ensuremath{r}\xspace}
\newcommand{\sigstrHat}{\ensuremath{\widehat{\sigstr}}\xspace}
\newcommand{\sigstrHatDel}{\ensuremath{\Delta\widehat{\sigstr}}\xspace}
\newcommand{\nuisan}{\ensuremath{\theta}\xspace}
\newcommand{\nuisanHat}{\ensuremath{\widehat{\nuisan}}\xspace}
\newcommand{\nuisanHatR}{\ensuremath{\nuisanHat_\sigstr}\xspace}
\newcommand{\nuisanZero}{\ensuremath{\nuisan_0}\xspace}
\newcommand{\nuisanDel}{\ensuremath{\Delta\nuisan}\xspace}
\newcommand{\nuisanPull}{\ensuremath{(\nuisanHat-\nuisanZero)/\nuisanDel}\xspace}
\newcommand{\testst}{\ensuremath{q(\sigstr)}\xspace}

\cmsNoteHeader{TOP-22-013}
\title{Observation of four top quark production in proton-proton collisions at \texorpdfstring{\sqrts}{sqrt(s) = 13 TeV}}

\date{\today}

\abstract{The observation of the production of four top quarks in proton-proton collisions is reported, based on a data sample collected by the CMS experiment at a center-of-mass energy of 13\TeV in 2016--2018 at the CERN LHC and corresponding to an integrated luminosity of 138\fbinv. Events with two same-sign, three, or four charged leptons (electrons and muons) and additional jets are analyzed. Compared to previous results in these channels, updated identification techniques for charged leptons and jets originating from the hadronization of \PQb quarks, as well as a revised multivariate analysis strategy to distinguish the signal process from the main backgrounds, lead to an improved expected signal significance of \sigexp standard deviations above the background-only hypothesis. Four top quark production is observed with a significance of \sigobs standard deviations, and its cross section is measured to be \xsecmeas, in agreement with the available standard model predictions.
}

\hypersetup{%
    pdfauthor={CMS Collaboration},%
    pdftitle={Observation of four top quark production in proton-proton collisions at sqrt(s) = 13 TeV},%
    pdfsubject={CMS},%
    pdfkeywords={CMS, top quark},%
}

\maketitle

\section{Introduction}

\begin{figure}[!ht]
\centering
\includegraphics[width=0.22\textwidth]{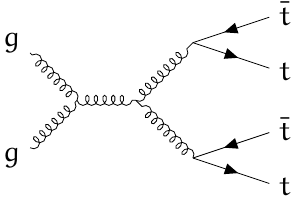}%
\cmsSpace{0.05\textwidth}%
\includegraphics[width=0.22\textwidth]{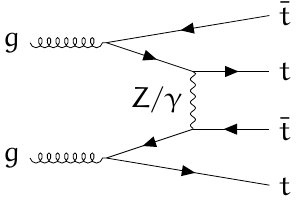}%
\cmsSpace{0.05\textwidth}%
\includegraphics[width=0.22\textwidth]{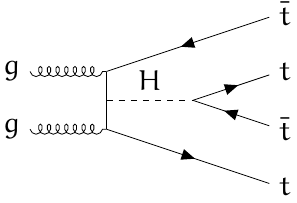}%
\caption{%
    Examples of Feynman diagrams that provide important contributions to \tttt production.
    The first diagram (\cmsUpperLeft) involves only the strong interaction.
    The other two diagrams involve both strong and electroweak interactions with the exchange of a virtual \PZ boson or photon (\cmsUpperRight), or a virtual Higgs boson (\cmsLower).
}
\label{fig:feynman}
\end{figure}

Four top quark (\tttt) production in proton-proton (\pp) collisions is among the rarest standard model (SM) processes currently accessible at hadron colliders.
While the production occurs predominantly through the strong interaction~\cite{Bevilacqua:2012em, EXT:MG5aMCatNLO-2014, Maltoni:2015ena}, nonnegligible contributions arise also from electroweak (EW) processes~\cite{Frederix:2017wme, Jezo:2021smh, vanBeekveld:2022hty}.
Example leading-order (LO) Feynman diagrams are shown in Fig.~\ref{fig:feynman}.
The SM cross section is calculated at next-to-LO (NLO) in quantum chromodynamics (QCD) and EW theory, including soft-gluon emission corrections at next-to-leading logarithmic accuracy, to be \xsecpred at \sqrts~\cite{vanBeekveld:2022hty}.
The quoted uncertainty is from scale variations and the parton distribution functions (PDFs).

Due to EW contributions involving the exchange of a virtual Higgs boson (\PH), as shown in Fig.~\ref{fig:feynman} (\cmsLower), the measurement of the \tttt production cross section provides a way to measure the top quark Yukawa coupling~\cite{Cao:2016wib, Cao:2019ygh}, complementary to its extraction from measurements of Higgs boson production in association with a top quark pair (\ttbar)~\cite{CMS:HIG-19-008} or \ttbar cross section measurements~\cite{CMS:TOP-17-004, CMS:TOP-19-008}.
The \tttt production process is also of interest as a probe for new physics.
Many models of physics beyond the SM (BSM) introduce an extension of the Higgs sector, resulting in additional scalar particles that have Yukawa-like interactions with top quarks.
Such new particles would enhance the \tttt production cross section~\cite{Dicus:1994bm, Craig:2015jba, Craig:2016ygr, Anisha:2023xmh}.
Supersymmetric extensions of the SM predict new heavy, strongly interacting particles that can decay to top quarks, leading to \tttt production via intermediate supersymmetric particles~\cite{Nilles:1983ge, Farrar:1978xj, Toharia:2005gm, Plehn:2008ae, Calvet:2012rk, Beck:2015cga, Darme:2018dvz}.
Other BSM scenarios are predicted to affect \tttt production as well~\cite{Kumar:2009vs, Cacciapaglia:2015eqa, Ducu:2015fda}, and model-independent evaluations of modifications to the SM in an effective field theory setup have shown the importance of \tttt production measurements to constrain these~\cite{Degrande:2010kt, Zhang:2017mls, Hartland:2019bjb, Englert:2019zmt, Banelli:2020iau, Darme:2021gtt, Ethier:2021bye, Aoude:2022deh}.

The ATLAS~\cite{ATLAS:Detector-2008} and CMS~\cite{CMS:Detector-2008} Collaborations at the CERN LHC performed searches for \tttt production using the \pp collision data recorded between 2015 and 2018, corresponding to an integrated luminosity of about 140\fbinv and covering the decay channels with zero to four electrons and/or muons (in the following referred to as ``leptons'')~\cite{CMS:SUS-16-035, CMS:TOP-17-009, ATLAS:2018alq, ATLAS:2018kxv, CMS:TOP-17-019, CMS:TOP-18-003, ATLAS:2020hpj, ATLAS:2021kqb, CMS:TOP-21-005, Blekman:2022jag}.
The dilepton channel is further divided into events where the charge of the two leptons has the same or opposite sign (in the following referred to as same- or opposite-sign lepton pairs, respectively).
So far, these searches found evidence for \tttt production with a significance of more than three standard deviations (SDs) from the background-only hypothesis~\cite{ATLAS:2020hpj, ATLAS:2021kqb, CMS:TOP-21-005}.
However, the observation level of five SDs~\cite{Lyons:2018gtc} was not reached.
In both experiments, the \tttt signal was measured with a significance larger than the expected one, and with a measured cross section higher than the SM prediction~\cite{ATLAS:2020hpj, ATLAS:2021kqb, CMS:TOP-21-005}, which could indicate an enhancement caused by the presence of BSM physics, but might also be explained as statistical fluctuations or point towards a systematic bias that is not properly understood.
Thus, it is of crucial importance to further increase the sensitivity of the \tttt measurements.

In this Letter, we present a search for \tttt production in events with two same-sign, three, or four leptons, using \pp collision data recorded by the CMS experiment in 2016--2018 and corresponding to an integrated luminosity of 138\fbinv.
This measurement supersedes the results from Ref.~\cite{CMS:TOP-18-003} that analyzed events with two same-sign or at least three leptons selected from the same data set and found 2.6 (2.7) SDs of observed (expected) significance for \tttt production.
Notable improvements, discussed later in this Letter, are achieved in the lepton identification and the tagging of jets originating from the hadronization of bottom (\PQb) quarks, as well as from a revised analysis strategy for the discrimination between signal and background processes based on the application of machine learning techniques.
The \tttt production cross section \sigtttt is extracted with a profile likelihood fit to distributions that provide optimal signal-to-background discrimination.
The backgrounds of \ttbar production in association with a \PW or \PZ boson (\ttW and \ttZ, respectively) are estimated with free normalization parameters in the fit.
The aforementioned improvements increase the sensitivity of the analysis and allow for the observation of the \tttt production process with a statistical significance above five SDs.
When this Letter was in the final stages of preparation, the ATLAS Collaboration also reported the observation of \tttt production~\cite{ATLAS:2023ajo} with an observed (expected) significance of 6.1 (4.3) SDs.

Tabulated results are provided in the HEPData record for this analysis~\cite{HEPData}.

\section{The CMS detector and event reconstruction}

The central feature of the CMS apparatus is a superconducting solenoid of 6\unit{m} internal diameter, providing a magnetic field of 3.8\unit{T}.
Within the solenoid volume are a silicon pixel and strip tracker, a lead tungstate crystal electromagnetic calorimeter (ECAL), and a brass and scintillator hadron calorimeter, each composed of a barrel and two endcap sections.
Forward calorimeters extend the pseudorapidity (\sigeta) coverage provided by the barrel and endcap detectors.
Muons are measured in gas-ionization detectors embedded in the steel flux-return yoke outside the solenoid.
A more detailed description of the CMS detector, together with a definition of the coordinate system used and the relevant kinematic variables, can be found in Ref.~\cite{CMS:Detector-2008}.

Events of interest are selected using a two-tiered trigger system.
The first level, composed of custom hardware processors, uses information from the calorimeters and muon detectors to select events at a rate of around 100\unit{kHz} within a fixed latency of about 4\mus~\cite{CMS:TRG-17-001}.
The second level, known as the high-level trigger, consists of a farm of processors running a version of the full event reconstruction software optimized for fast processing, and reduces the event rate to around 1\unit{kHz} before data storage~\cite{CMS:TRG-12-001}.

A global particle-flow (PF) algorithm~\cite{CMS:PRF-14-001} aims to reconstruct and identify each individual particle in an event, with an optimized combination of information from the various elements of the CMS detector.
The primary vertex (PV) is taken to be the vertex corresponding to the hardest scattering in the event, as described in Section 9.4.1 of Ref.~\cite{CMS:TDR-15-02}.
The energy of photons is obtained from the ECAL measurement.
The energy of electrons is obtained from a combination of the electron momentum at the PV as determined by the tracker, the energy of the corresponding ECAL cluster, and the energy sum of all bremsstrahlung photons spatially compatible with originating from the electron track.
The energy of muons is obtained from the curvature of the corresponding track.
The energy of charged hadrons is determined from a combination of their momentum measured in the tracker and the matching calorimeter energy deposits, corrected for the response function of the calorimeters to hadronic showers.
Finally, the energy of neutral hadrons is obtained from the corresponding corrected calorimeter energies.

Hadronic jets are clustered from the PF objects using the infrared- and collinear-safe anti-\kt algorithm~\cite{EXT:AK-2008, EXT:FastJet-2012} with a distance parameter of 0.4.
Jet momentum is determined as the vector sum of all PF object momenta in the jet, and is found from simulation to be, on average, within 5--10\% of the true momentum over the entire transverse momentum (\pt) spectrum and detector acceptance.
Additional \pp interactions within the same or nearby bunch crossings (pileup) can contribute additional tracks and calorimetric energy depositions, increasing the apparent jet momentum.
To mitigate this effect, charged PF objects identified to be originating from pileup vertices are discarded and an offset correction is applied to correct for remaining contributions.
Jet energy corrections are derived from simulation studies so that the average measured energy of jets becomes identical to that of particle-level jets.
In situ measurements of the momentum balance in dijet, photon+jet, \PZ{+}jet, and multijet events are used to determine any residual differences between the jet energy scale (JES) in data and simulation, and appropriate corrections are made.
The jet energy resolution (JER) amounts typically to 15--20\% at 30\GeV, 10\% at 100\GeV, and 5\% at 1\TeV.
A smearing procedure is applied to match the JER in simulation to that in data~\cite{CMS:JME-13-004}.
Additional selection criteria are applied to each jet to remove jets potentially dominated by instrumental effects or reconstruction failures.
We retain jets for further analysis if they have $\pt>25\GeV$ and $\abseta<2.4$, and are separated by $\DR=\sqrt{\smash[b]{(\Deta)^2+(\Dphi)^2}}>0.4$ from any identified lepton, where \Deta and \Dphi are the \sigeta and azimuthal angle differences between the jet and lepton directions.

The missing transverse momentum vector \ptvecmiss is computed as the negative vector sum of the \ptvec of all the PF objects in an event, including unclustered energy from the PF objects not associated with any reconstructed lepton, photon, or jet~\cite{CMS:JME-17-001}.
Its magnitude is denoted as \ptmiss.
The \ptvecmiss is modified to account for corrections to the energy scale of the reconstructed jets in the event.
Anomalous high-\ptmiss events can be due to a variety of reconstruction failures, detector malfunctions, or noncollision backgrounds.
Such events are rejected by event filters that are designed to identify more than 85--90\% of the spurious high-\ptmiss events with a mistagging rate less than 0.1\%~\cite{CMS:JME-17-001}.

Jets originating from \PQb quarks are identified with the \DeepJet algorithm~\cite{CMS:BTV-16-002, EXT:DeepJet-2020, CMS:DP-2023-005}, with three defined working points (WPs) labeled ``loose'', ``medium'', and ``tight''.
The loose WP has a selection efficiency for \PQb quark jets of about 90\%, and a misidentification rate of 49 (18)\% for \PQc quark jets (light-quark and gluon jets evaluated together and referred to as ``light jets'' in the following).
Similarly, the efficiencies for \PQb quark, \PQc quark, and light jets are about 76, 17, and 3\% for the medium WP, and 59, 3, and 0.3\% for the tight WP.
Compared to the \textsc{DeepCSV} algorithm used in Ref.~\cite{CMS:TOP-18-003}, the \PQb quark jet selection efficiency at the same light jet misidentification rate is 5--25\% higher with the \DeepJet algorithm~\cite{CMS:DP-2023-005}.
Unless specified otherwise, the term ``\PQb jets'' is used to refer to jets that pass the loose WP requirements.

\section{Simulated event samples}
\label{sec:simulation}

Simulated event samples of the signal and background processes are generated with Monte Carlo generators and used to determine the \tttt signal acceptance, estimate most background contributions, and provide training data for the machine-learning discriminants.
Different matrix-element (ME) generators are used to simulate different processes as described below, and in all cases the NNPDF3.1NNLO~\cite{EXT:NNPDF-2017} PDFs are used.
The simulation of parton showering, hadronization, and underlying event is performed with the \PYTHIA v8.230 program~\cite{EXT:Pythia-2015} using the CP5 tune~\cite{CMS:GEN-17-001}.
Pileup collisions are overlaid to each simulated event, and the generated distribution of the number of events per bunch crossing is matched to that observed in data.
All simulated events are processed with a full simulation of the CMS detector based on the \GEANTfour toolkit~\cite{EXT:GEANT4-2002}.

The \tttt signal process is simulated at NLO in QCD with the \MGvATNLO v2.6.5 program~\cite{EXT:MG5aMCatNLO-2014}, where the decay of the top quarks is simulated with \textsc{madspin}~\cite{EXT:Madspin-2013}.
The background processes of \ttbar production in association with a \PZ (\ttZ) or a \PW (\ttW) boson are simulated as $\ttbar\Pell\Pell$ and $\ttbar\Pell\PGn$, respectively, at NLO in QCD with \MGvATNLO, and thus include nonresonant contributions, and the top quark decays are also simulated with \textsc{madspin}.
The \ttZ sample is generated for invariant masses of $\mll>10\GeV$.
The \ttW sample is generated for invariant masses of $m(\Pell\PGn)>30\GeV$, and up to one additional parton is included in the ME simulation.
To eliminate double counting of partons generated with \MGvATNLO and \PYTHIA in the \ttW sample, the FxFx matching scheme~\cite{EXT:FxFx-2012} is used, and more details of the specific settings are given in Ref.~\cite{Ferencz:2023fso}.
Three top quark production is simulated as $\ttbar\PQt\PAQq$, $\ttbar\PAQt\PQq$, $\ttbar\PQt\PWm$, and $\ttbar\PAQt\PWp$, labeled together as \ttt, at LO in QCD with \MGvATNLO using the five-flavor scheme.
The renormalization and factorization scales in the ME calculation are chosen dynamically as half the scalar sum of the transverse masses of all final-state particles for the \tttt, \ttZ, and \ttt samples, and dependent on the kinematic properties of the event following the FxFx merging prescription (as described in Section 2.2.3 of Ref.~\cite{EXT:FxFx-2012}) for the \ttW sample.

Further background samples simulated with \MGvATNLO at NLO in QCD are \ttbar production in association with photon or Higgs boson, single top quark production (\PZ boson associated and \WZ diboson associated), $\PZ/\PGg^\ast$ production in association with jets (referred to as \Zjets), and \WZ, $\PZ\PGg$, $\PW\PH$, and $\PZ\PH$ diboson production.
Simulated at LO in QCD with \MGvATNLO are furthermore \ttbar production in association with two bosons, single top quark production in association with a Higgs boson, and same-sign \WW diboson production.
Double counting of partons is eliminated with the FxFx (MLM~\cite{EXT:Merging-2008}) matching scheme for NLO (LO) samples.

The \POWHEG v2 program~\cite{EXT:Powheg-2004, EXT:Powheg-ttbar-2007, EXT:Powheg-2007, EXT:Powheg-singlet-2009, Nason:2009ai, EXT:Powheg-2010, Re:2010bp, EXT:Powheg-ggH-2011, Nason:2013ydw} is used for the simulation of \ttbar production, quark-antiquark-initiated \ZZ and opposite-sign \WW diboson production, and Higgs boson production via gluon and vector boson fusion at NLO in QCD.
In \POWHEG samples, the Higgs boson decay to four leptons is simulated with the \textsc{JHUgen} v5.2.5 program~\cite{Bolognesi:2012mm}.
The gluon-gluon-initiated \ZZ diboson production process is simulated with the \MCFM v7.0.1 generator~\cite{Campbell:1999ah, Campbell:2011bn, Campbell:2015qma} at LO in QCD.

\section{Lepton selection}

Electrons are reconstructed via a combination of tracker and ECAL measurements~\cite{CMS:EGM-17-001, CMS:DP-2020-021} in the range $\abseta<2.5$, and electrons in the barrel--endcap transition region $1.44<\abseta<1.57$ are removed because of nonoptimal electron reconstruction in this region.
The curvature of the electron track is evaluated with three different methods to estimate the electron charge, and we require all three charge evaluations to agree for each selected electron, thereby reducing the background from charge mismeasurements (referred to as ``charge misID'') by a factor of five with an efficiency of about 97\%~\cite{CMS:EGM-13-001}.
Muons are reconstructed in the range $\abseta<2.4$ by combining information from the tracker, the muon spectrometers, and the calorimeters in a global fit~\cite{CMS:MUO-16-001}.
The charge mismeasurements of muons are negligible~\cite{CMS:CFT-09-014, CMS:MUO-17-001}.

We select electrons and muons with $\pt>10\GeV$ and require that they are compatible with originating from the PV and fulfill a loose set of identification (ID) criteria~\cite{CMS:EGM-17-001, CMS:MUO-16-001}.
The relative isolation \Irel of a lepton is defined as the scalar \pt sum of all particles within a certain distance \DR\ around the lepton, divided by the lepton \pt.
We use a \pt-dependent distance~\cite{Rehermann:2010vq} of $\DR<0.2$ for leptons with $\pt<50\GeV$, $\DR<10\GeV/\pt$ for $50<\pt<200\GeV$, and $\DR<0.05$ for $\pt>200\GeV$, and require all leptons to pass a loose requirement of $\Irel<0.4$.

Signal events are characterized by the presence of leptons originating directly from the prompt decay of top quarks or massive bosons (referred to as ``prompt leptons'').
Background contributions with ``nonprompt leptons'' arise from events with genuine leptons produced in hadron decays and photon conversions, as well as from events with jet constituents misidentified as leptons.
To distinguish prompt and nonprompt leptons, we employ gradient boosted decision trees (BDTs) trained with the \textsc{XGBoost} program~\cite{Chen:2016btl} separately for electrons and muons, following the methods developed for various CMS measurements and searches with multilepton signatures~\cite{CMS:HIG-17-018, CMS:TOP-18-008, CMS:HIG-19-008, CMS:SUS-19-012, CMS:SMP-20-012, CMS:TOP-20-010}.
Simulated \tttt, \ttW, \ttZ, \ttH, and \tZq (\ttbar) samples are used to provide the prompt-lepton (nonprompt-lepton) training data.
The input variables for the BDTs are the lepton \pt and \abseta, isolation information, variables quantifying the consistency of the lepton track with the PV, and properties of the nearest jet defined as the jet that includes the PF particle corresponding to the reconstructed lepton.
For electrons, the multivariate discriminant from Ref.~\cite{CMS:EGM-17-001} is used as additional input.
For muons, an additional input is the segment compatibility defined in Ref.~\cite{CMS:MUO-16-001}.
Besides the lepton \pt, the most important input variables are the ratio of the lepton \pt to the nearest jet \pt and the \DeepJet score of the nearest jet, and also the multivariate discriminant in the case of the electron BDT.
A dedicated study of the BDT strategy applied here with the full list of input variables is presented in Ref.~\cite{CMS:MUO-22-001} for the case of muons.

\ifthenelse{\boolean{cms@external}}{\begin{figure*}[!t]}{\begin{figure*}[!b]}
\centering
\includegraphics[width=0.48\textwidth]{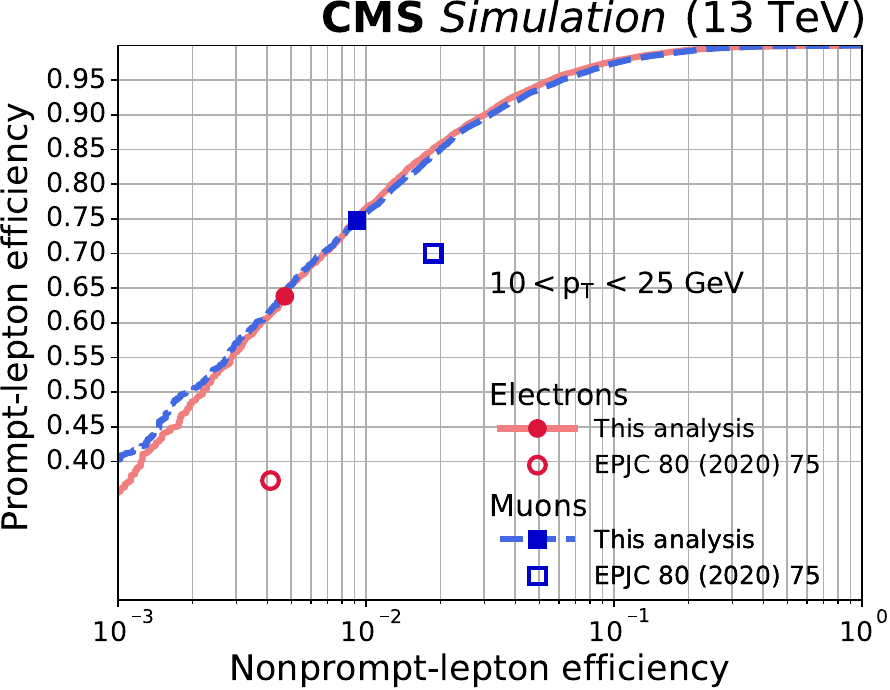}%
\hfill%
\includegraphics[width=0.48\textwidth]{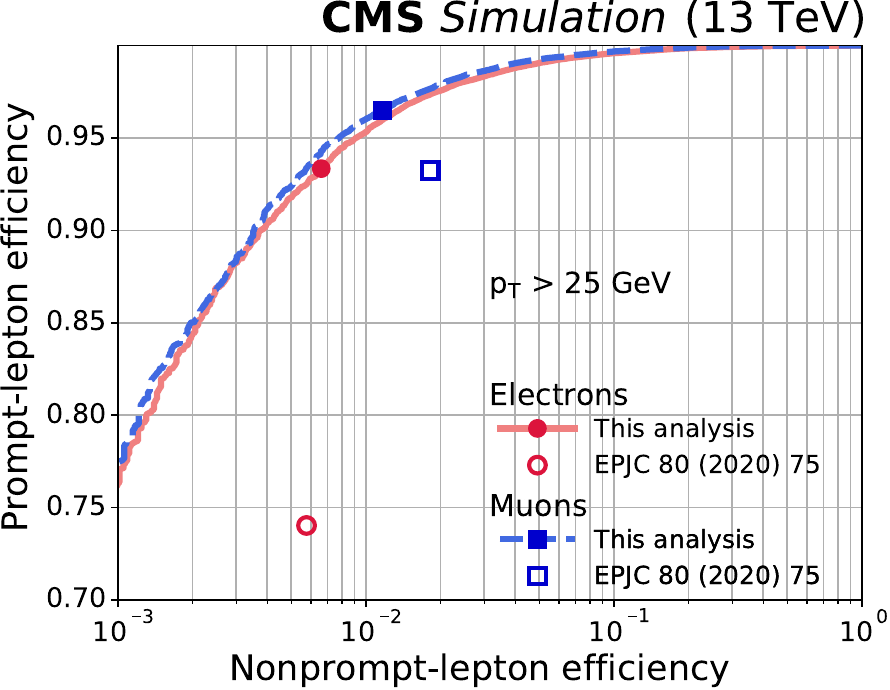}%
\caption{%
    Efficiency of selecting prompt leptons as a function of the misidentification probability for nonprompt leptons evaluated in simulated \ttbar events for the electron (red solid line) and muon (blue dashed line) ID BDT, shown for leptons with $10<\pt<25\GeV$ (left) and $\pt>25\GeV$ (right).
    Indicated with filled markers are the efficiencies for the ID criteria applied in this measurement and with empty markers those for the ID criteria applied in Ref.~\cite{CMS:TOP-18-003}, where red circles and blue squares are used for electron and muon criteria, respectively.
}
\label{fig:leptonid}
\end{figure*}

We define two sets of lepton ID criteria labeled ``loose'' and ``tight''.
Electrons and muons pass the tight ID if their BDT discriminant value is above a certain threshold.
The efficiencies of this selection are shown in Fig.~\ref{fig:leptonid}, and compared to the efficiencies of the ID criteria used in Ref.~\cite{CMS:TOP-18-003}.
To compensate for larger background contributions from nonprompt electrons compared to those from nonprompt muons, the threshold for tight electrons is set at a lower misidentification probability for nonprompt leptons, resulting also in a smaller prompt-lepton efficiency.
The loose ID is defined by requiring leptons to either pass the tight ID or a set of thresholds on the \pt ratio and the nearest jet \DeepJet score.
The thresholds are chosen such that the probabilities for nonprompt leptons to pass the loose ID are similar in simulated \ttbar and \Zjets events, ensuring that the prediction of backgrounds with nonprompt leptons described in Section~\ref{sec:background} is valid for all sources of nonprompt leptons.

\section{Event selection and search strategy}

The analyzed event sample is collected with a combination of triggers that require the presence of one, two, or three leptons.
Events must contain between two and four loose leptons, with $\pt>25$ and 20\GeV for the highest \pt (leading) and second-highest \pt (subleading) lepton, and at least two jets, of which at least one is identified as \PQb jet.
Events with two (three and four) leptons are removed if any lepton pair has an invariant mass below 20 (12)\GeV, to reduce backgrounds from leptonic decays of low-mass resonances.
Signal regions (SRs) and control regions (CRs) are defined using events in which all leptons pass the tight ID criteria, whereas events with at least one loose but not tight lepton are used as a sideband for the nonprompt-lepton background estimation.
In events with two leptons, the SRs and CRs additionally require both leptons to have the same sign, and events with opposite-sign leptons are used as a sideband for the estimation of the charge-misID background.
In events with four leptons, we additionally require the sum of the lepton charges to be zero.
No requirement on the lepton charges is imposed in events with three leptons.
For the SR and CR definitions, the number of jets and \PQb jets (\Njets and \Nbjets), the scalar \pt sum of all jets (\HT), and the invariant mass \mll of opposite-sign same-flavor (OSSF) lepton pairs are used.
A schematic representation of the SR and CR definitions is shown in Fig.~\ref{fig:eventselection}.

\begin{figure*}[!ht]
\centering
\includegraphics[width=\textwidth]{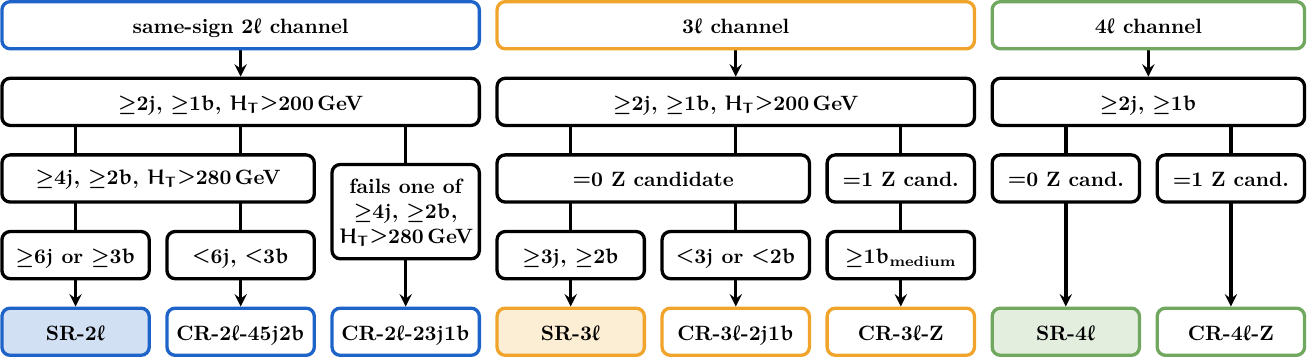}
\caption{%
    Schematic representation of the event selection and categorization.
}
\label{fig:eventselection}
\end{figure*}

For the events with two same-sign leptons (\twol channel), the \SRtwo is defined by $\Njets\geq4$, $\Nbjets\geq2$, and $\HT>280\GeV$, and additionally either $\Njets\geq6$ or $\Nbjets\geq3$.
The \CRtwoHigh, enriched in \ttW production, comprises all events with $4\leq\Njets\leq5$, $\Nbjets=2$, and $\HT>280\GeV$.
The \CRtwoLow, used to constrain both \ttW production and nonprompt-lepton backgrounds, is defined by all events with $\HT>200\GeV$ that fail exactly one of the requirements $\Njets\geq4$, $\Nbjets\geq2$, or $\HT>280\GeV$.

The \SRthree for events with three leptons (\threel channel) is defined by $\Njets\geq3$, $\Nbjets\geq2$, $\HT>200\GeV$, and the requirement that there is no OSSF lepton pair with $\abs{\mll-\mZ}<15\GeV$ (referred to as ``\PZ candidate'').
The last requirement, which uses the world-average \PZ boson mass~\cite{EXT:PDG-2022}, rejects events consistent with leptonically decaying \PZ bosons.
The \CRthreeLow, used to constrain nonprompt-lepton backgrounds, consists of events that pass the \SRthree requirements except that they have $\Njets=2$ or $\Nbjets=1$.
To constrain \ttZ production, the \CRthreeZ is defined by requiring $\HT>200\GeV$, the presence of a \PZ candidate, and that at least one jet passes the medium \DeepJet WP.

No additional jet requirements are imposed on events with four leptons (\fourl channel).
Events that have no \PZ candidate form the \SRfour, while events with exactly one \PZ candidate form the \CRfourZ enriched in \ttZ production.

To enhance the separation of signal events and those from different background processes, we employ multiclassification BDTs trained with the \TMVA program~\cite{EXT:TMVA-2007}.
The BDTs provide output scores for three classes of events:\ the \tttt signal, associated \ttbar production with a heavy boson (\ttZ, \ttW, and \ttH, referred to as \ttX), and \ttbar production, which is the dominant contribution to the nonprompt-lepton and charge-misID backgrounds.
Simulated event samples of \tttt, \ttZ, \ttW, \ttH, and \ttbar production are used in the training.
Kinematic differences and different background compositions motivate separate BDT trainings per decay channel, but sufficient simulated events in the training samples are required as well.
Since the size of the available \fourl samples is limited, we train one BDT for the combined \threel{+}\fourl channel and a second one for the \twol channel.
Only events that pass the \SRtwo (\SRthree or \SRfour) selection requirements are used in the \twol (\threel{+}\fourl) BDT training.

The variables used in the BDT training are listed in Table~\ref{tab:inputeventbdt}.
From an initially wider set of input variables, we optimized the selection of variables by evaluating how much the removal of a specific observable from the training reduces the BDT performance, simultaneously taking into account the accuracy of all three classes, and retaining only those variables with a noticeable impact on the performance.
This was done separately for the \twol and \threel{+}\fourl channels, resulting in differences in the list of input variables.
The highest ranked variables are \DeepJet scores of selected jets, angular separations between leptons and \PQb jets, and invariant masses of two- and three-jet systems consistent with the decay of a hadronically decaying \PW boson or top quark.
The modeling of each input variable in the simulated samples is validated in data.
The BDT scores can be interpreted as measures of how likely an event is to originate from the corresponding classes, and good agreement between simulation and data is found for the shapes of their distributions in the CRs.

\begin{table*}[!p]
\centering
\renewcommand{\arraystretch}{1.25}
\topcaption{%
    List of the input variables to the event multiclassification BDTs.
    The last two columns indicate the importance rank of the observables in the \twol and \threel{+}\fourl BDT trainings, respectively, as evaluated with \TMVA~\cite{EXT:TMVA-2007}, and a dash indicates that the observable is not used in that training.
    The \mTii variable, defined in Refs.~\cite{Lester:1999tx, Lester:2011nj}, is constructed from \ptvecmiss and two four-momenta of the particles or particle systems specified in the table.
}
\cmsTable{\begin{tabular}{cp{0.85\textwidth}cc}
    Symbol & Definition & \twol & \threel{+}\fourl \\
    \hline
    $\max_2\DpJ$ & Second-highest \DeepJet score of any jet & 1 & 1 \\
    $\max_3\DpJ$ & Third-highest \DeepJet score of any jet & 8 & 15 \\
    $\max_4\DpJ$ & Fourth-highest \DeepJet score of any jet & 29 & \NA \\
    $\DpJ(\PQj_1)$ & \DeepJet score of leading jet & 9 & 7 \\
    $\DpJ(\PQj_2)$ & \DeepJet score of subleading jet & 12 & 11 \\
    $\DpJ(\PQj_3)$ & \DeepJet score of jet with third-highest \pt & 16 & 18 \\
    $\DpJ(\PQj_4)$ & \DeepJet score of jet with fourth-highest \pt & \NA & 22 \\[\cmsTabSkip]
    $\DR(\Pell_1,\Pell_2)$ & \DR\ between leading and subleading lepton & 2 & 4 \\
    $\Dphi(\Pell_1,\Pell_2)$ & \Dphi between leading and subleading lepton & 5 & \NA \\
    $\min\DR(\PQb,\PQb)$ & Smallest \DR\ between any two \PQb jets & 3 & 3 \\
    $\min\DR(\Pell,\PQb)$ & Smallest \DR\ between any lepton and \PQb jet & 6 & \NA \\
    $\min_2\DR(\Pell,\PQb)$ & Second-smallest \DR\ between any lepton and \PQb jet & 4 & 6 \\[\cmsTabSkip]
    \mtop1 & Invariant mass of any three jets, of which one is identified as \PQb jet, that is closest to the top quark mass & 7 & 2 \\
    \mW1 & Invariant mass of any two jets used for \mtop1 that is closest to the \PW boson mass & 10 & 5 \\
    \mtop2 & Invariant mass of any three jets, of which one is identified as \PQb jet and none of which is used for \mtop1, that is closest to the top quark mass & 23 & \NA \\
    \mW2 & Invariant mass of any two jets used for \mtop2 that is closest to the \PW boson mass & 28 & \NA \\[\cmsTabSkip]
    \HT & Scalar \pt sum of all jets & 14 & 8 \\
    \ptmiss & Missing transverse momentum & 20 & 13 \\
    \Njets & Number of jets & 24 & 23 \\
    $\Nbjets^{\text{tight}}$ & Number of jets passing the tight \DeepJet WP & 27 & 24 \\
    $\Nbjets^{\text{medium}}$ & Number of jets passing the medium \DeepJet WP & 30 & \NA \\[\cmsTabSkip]
    $\pt(\Pell_1)$ & Highest \pt of any lepton & 19 & 14 \\
    $\pt(\Pell_2)$ & Second-highest \pt of any lepton & 15 & 17 \\
    $\pt(\Pell_3)$ & Third-highest \pt of any lepton & \NA & 9 \\[\cmsTabSkip]
    $\pt(\PQj_1)$ & Highest \pt of any jet & 18 & 12 \\
    $\pt(\PQj_2)$ & Second-highest \pt of any jet & 22 & \NA \\
    $\pt(\PQj_3)$ & Third-highest \pt of any jet & \NA & 10 \\
    $\pt(\PQj_4)$ & Fourth-highest \pt of any jet & 11 & \NA \\
    $\pt(\PQj_5)$ & Fifth-highest \pt of any jet & 13 & 21 \\[\cmsTabSkip]
    $\mT(\Pell_1)$ & Transverse mass of leading lepton and \ptmiss & 17 & 19 \\
    $\mT(\Pell_2)$ & Transverse mass of subleading lepton and \ptmiss & 21 & 16 \\
    $\mTii(\Pell)$ & \mTii variable constructed from leading and subleading leptons & \NA & 20 \\
    $\mTii(\PQb)$ & \mTii variable constructed from leading and subleading \PQb jets & 25 & \NA \\
    $\mTii(\Pell{+}\PQb)$ & \mTii variable constructed from two lepton+jet systems built with leading two leptons and leading two \PQb jets & 26 & \NA \\
\end{tabular}}
\label{tab:inputeventbdt}
\end{table*}

Events selected in the SRs are further split according to the class that yields the highest BDT score, resulting in ``\tttt-like'', ``\ttX-like'', and ``\ttbar-like'' SR classes.
These SR classes, together with the CRs, are separate inputs to the fit for the cross section extraction.
Due to the low number of events, the \ttbar class is added to the \ttX class in \SRfour.

The approach followed here is different from the BDT approach of Ref.~\cite{CMS:TOP-18-003} in several aspects.
We train separate BDTs for the \twol and \threel{+}\fourl channels, with an extended set of input variables separately optimized for the two channels.
Especially the usage of \DeepJet discriminants of several jets and \DR\ information between leptons and \PQb jets, which were not used in Ref.~\cite{CMS:TOP-18-003}, improves the sensitivity of our BDTs, as indicated by the high importance ranking of these observables shown in Table~\ref{tab:inputeventbdt}.
The performance also benefits from looser requirements in the initial event selection, allowing an improvement of the BDTs with respect to the sensitivity achieved by a series of tight one-dimensional selections.
Finally, the setup as a multiclassifier that simultaneously separates two characteristically different background classes facilitates a good separation of processes, and the use of all output classes in the fit improves the constraints on the main background contributions.

\section{Background estimation}
\label{sec:background}

Background contributions to the SRs are categorized into processes in which all selected leptons are prompt (prompt background) and processes with at least one nonprompt lepton (nonprompt background).
In the \twol channel, we additionally separate prompt-background contributions from the charge-misID backgrounds.
Prompt backgrounds are estimated from the simulated samples described in Section~\ref{sec:simulation}, while nonprompt and charge-misID backgrounds are estimated from sideband samples in data.
All predictions from simulation are normalized according to the measured integrated luminosity and state-of-the-art cross section predictions.

Prompt-background contributions with top quarks are grouped into \ttX, \ttt, and other associated top quark production processes labeled ``{\otherT}''.
Contributions from \ttbar production, including \ttbb and other contributions with additional jets, do not contribute to the prompt background as they either need to have a charge-misID lepton or an additional nonprompt lepton to pass the event selection.
Instead, these processes are the dominant contributions to the nonprompt and charge-misID backgrounds.
Prompt-background contributions without top quarks are mainly from the production of two or three heavy vector bosons, grouped together as ``\VVV'', where the leptons are from leptonic boson decays and the jets originate mostly from additional QCD radiation.
Contributions with photon conversions, both in processes with top quarks and only vector bosons, are grouped together as ``\Xgamma''.

The \ttW background is relevant in the \twol and \threel channels, \ttZ in the \threel and \fourl channels with smaller contributions in the \twol channel, and \ttH in all channels.
The \ttW prediction is normalized to a cross section of \ttwpred~\cite{Frederix:2021agh}, \ttZ to \ttzpred~\cite{Kulesza:2020nfh}, and \ttH to $504\pm39\fb$~\cite{Kulesza:2020nfh}.
The quoted uncertainties are from scale variations and PDF uncertainties evaluated in the SM cross section calculations, and are different from the normalization uncertainties we consider in our measurement and discuss in Section~\ref{sec:systematics}.
The contribution from \ttt production to the SR \tttt classes is predicted to be less than 9\% of the \tttt yield, but can be important since the kinematic properties are similar to the \tttt signal process.
We normalize the \ttt contribution to a cross section of $2.0\pm0.3\fb$, calculated at NLO in QCD and including LO EW corrections~\cite{Durieux:2023ttt}.
The contributions in \otherT are small, with the largest remaining contribution being \tZq production, normalized to a cross section of $94.2\pm3.1\fb$ for simulated events with $\mll>30\GeV$~\cite{CMS:TOP-16-020}.
We show the \ttt and \otherT contributions separately in the figures for the SR \tttt classes, but merge \ttt into \otherT in all other cases.

The main background contributions in the \VVV group are \WZ production in the \twol and \threel channels, and \ZZ production in the \fourl channel.
The number of additional jets in the simulated \WZ and \ZZ samples commonly used in measurements of top quark processes is known to be underestimated~\cite{CMS:TOP-18-008, CMS:TOP-18-009, CMS:HIG-19-008} and thus studied in two data validation regions.
A sample of \WZ candidate events in data is selected by requiring three leptons, one \PZ candidate, a three-lepton invariant mass \mthreel with $\abs{\mthreel-\mZ}>15\GeV$, $\ptmiss>30\GeV$, and no \PQb jets.
More than 70\% of the selected events originate from \WZ production, with smaller contributions from nonprompt leptons and \ZZ production.
A sample of \ZZ candidate events in data is selected by requiring four leptons that form two \PZ candidates, and contributions from other processes to this selection are found to be negligible.
A disagreement between data and prediction is observed in the jet multiplicity distribution in both samples, and scale factors are derived per \Njets bin that increase the contribution of \WZ (\ZZ) production by factors of up to 2.7 (1.8) at 6 (3) or more jets.

\begin{figure*}[!tb]
\centering
\includegraphics[width=0.32\textwidth]{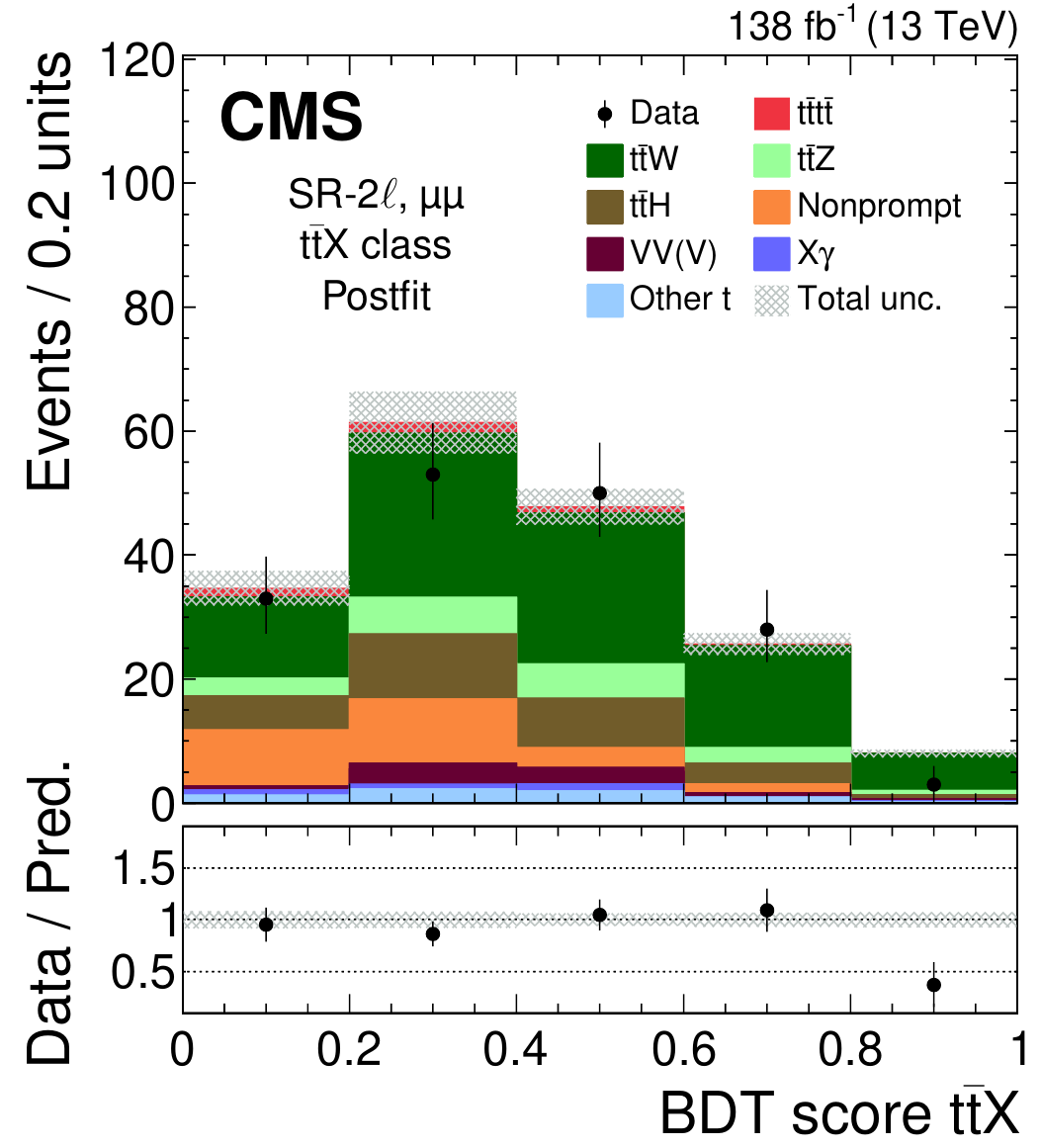}%
\hfill%
\includegraphics[width=0.32\textwidth]{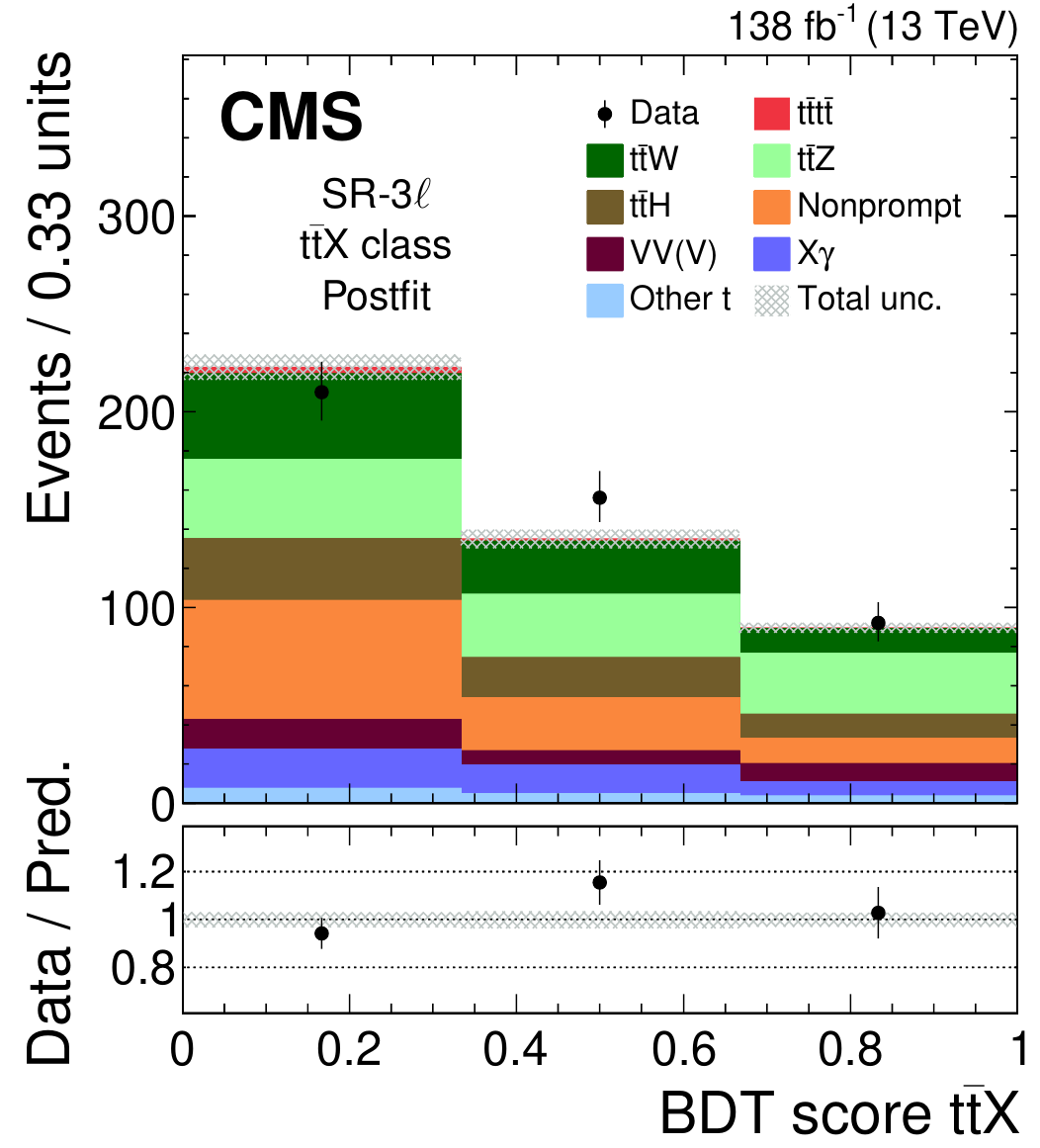}%
\hfill%
\includegraphics[width=0.32\textwidth]{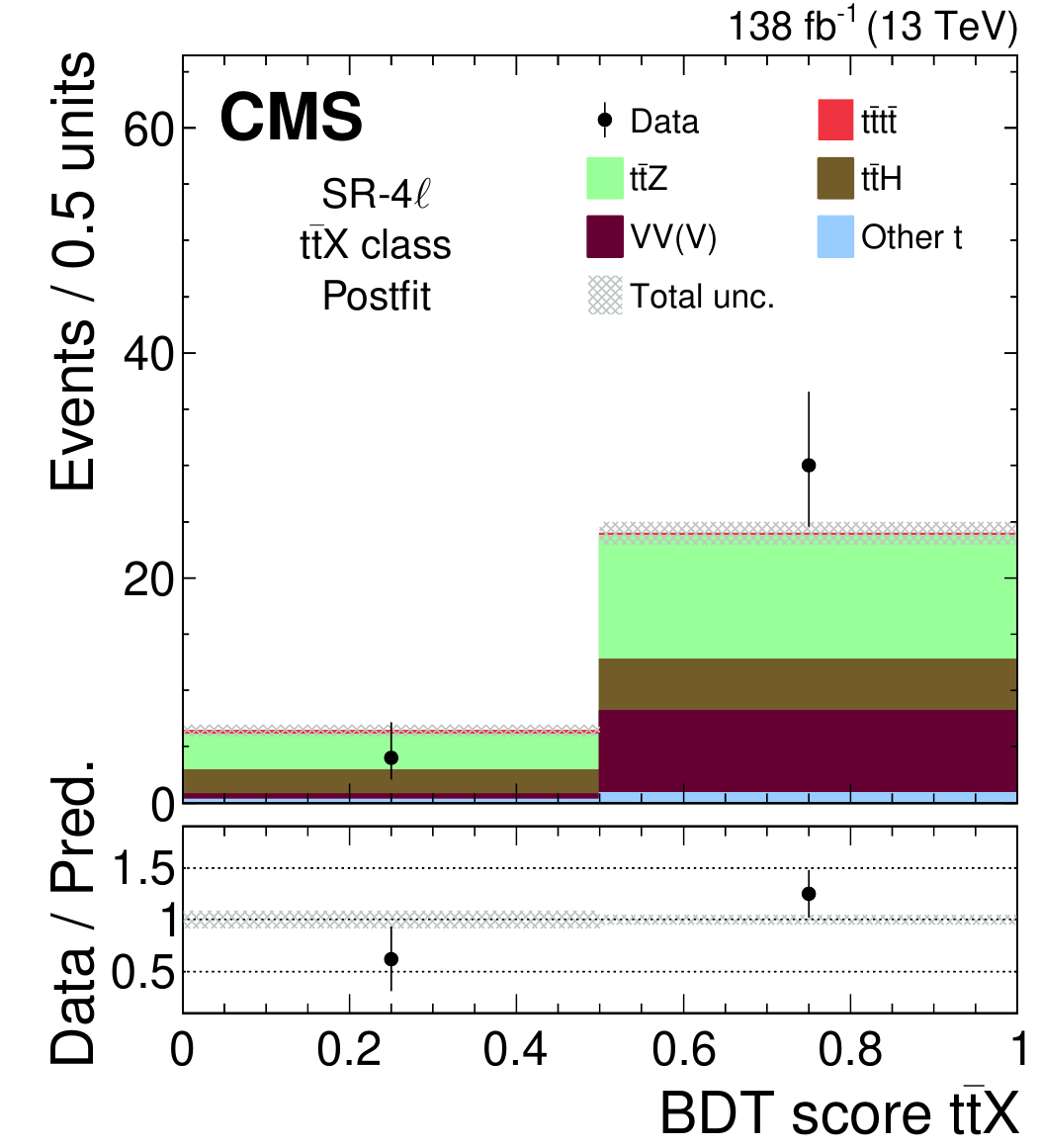}%
\caption{%
    Comparison of the number of observed (points) and predicted (colored histograms) events in the BDT score \ttX in the \ttX classes of \SRtwo in the \mumu category (left), of \SRthree (middle), and of \SRfour (right).
    The vertical bars on the points represent the statistical uncertainties in the data, and the hatched bands the total uncertainty in the predictions.
    The signal and background yields are shown with their best fit normalizations from the simultaneous fit to the data (``postfit'').
}
\label{fig:SR-ttX}
\end{figure*}

The contributions from the dominant prompt-background processes are treated separately in the fit.
Effective constraints are obtained from the inclusion of the SR \ttX classes, as well as the \CRthreeZ and \CRfourZ, in the cross section extraction.
Distributions of the BDT score \ttX in the SR \ttX classes are shown in Fig.~\ref{fig:SR-ttX} for the \mumu category of \SRtwo, \SRthree, and \SRfour.
Jet and \PQb jet multiplicity distributions in the \CRthreeZ and \CRfourZ are shown in Fig.~\ref{fig:CR-Z}.
Due to the high purity of these regions in \ttW and \ttZ production, we can extract normalization parameters for these processes from the fit without prior constraints.

\begin{figure*}[!tb]
\centering
\includegraphics[width=0.32\textwidth]{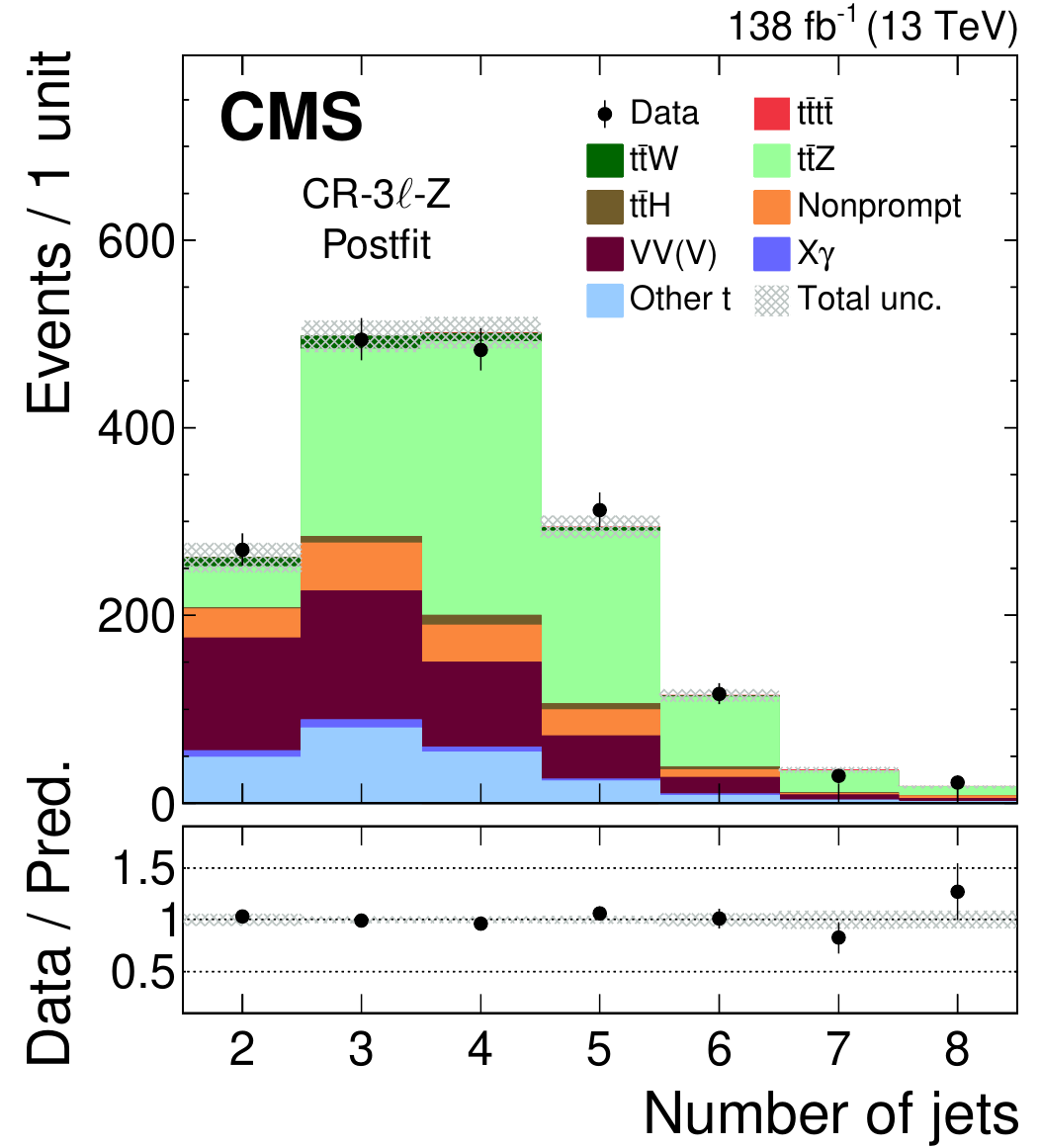}%
\hfill%
\includegraphics[width=0.32\textwidth]{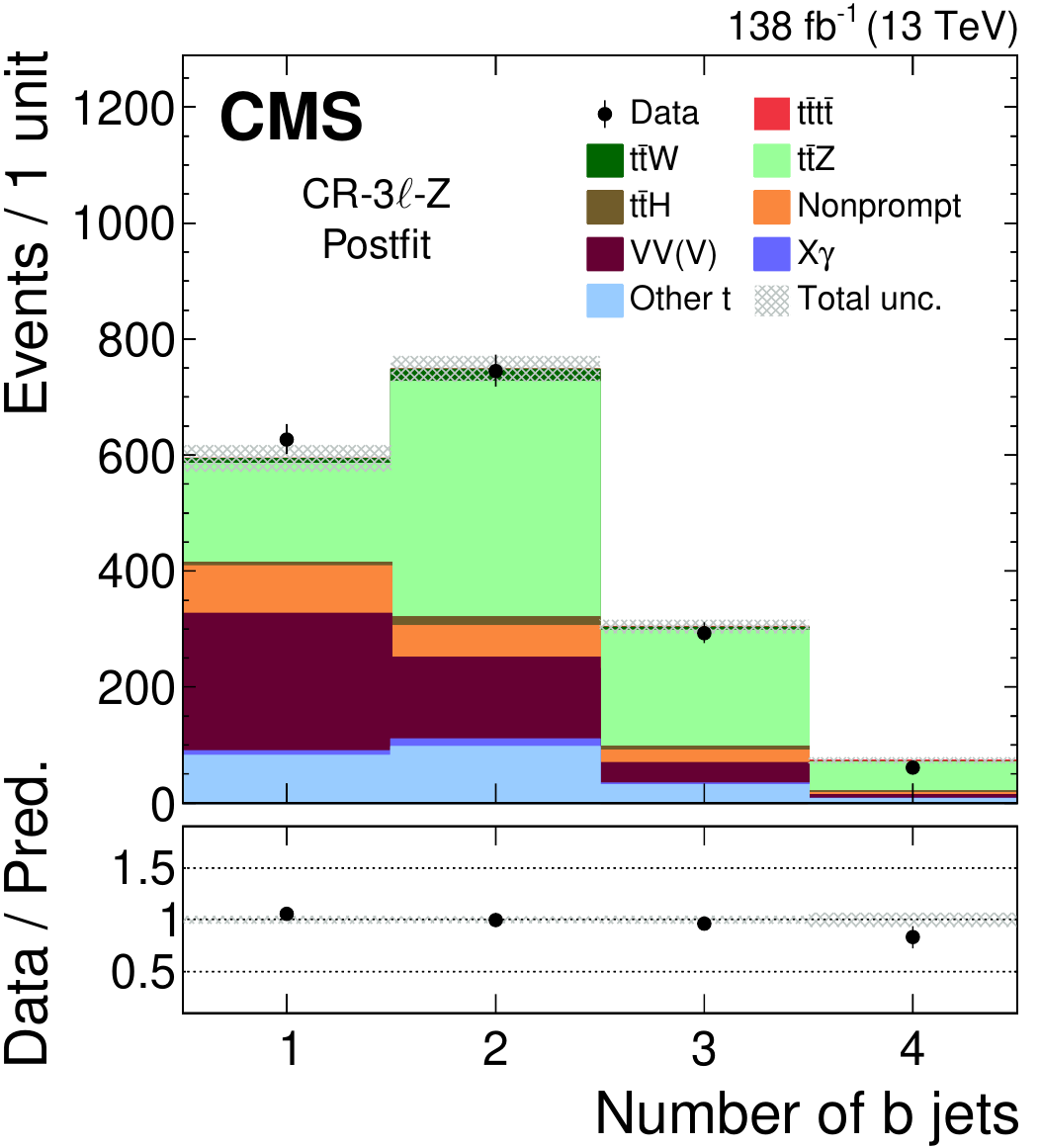}%
\hfill%
\includegraphics[width=0.32\textwidth]{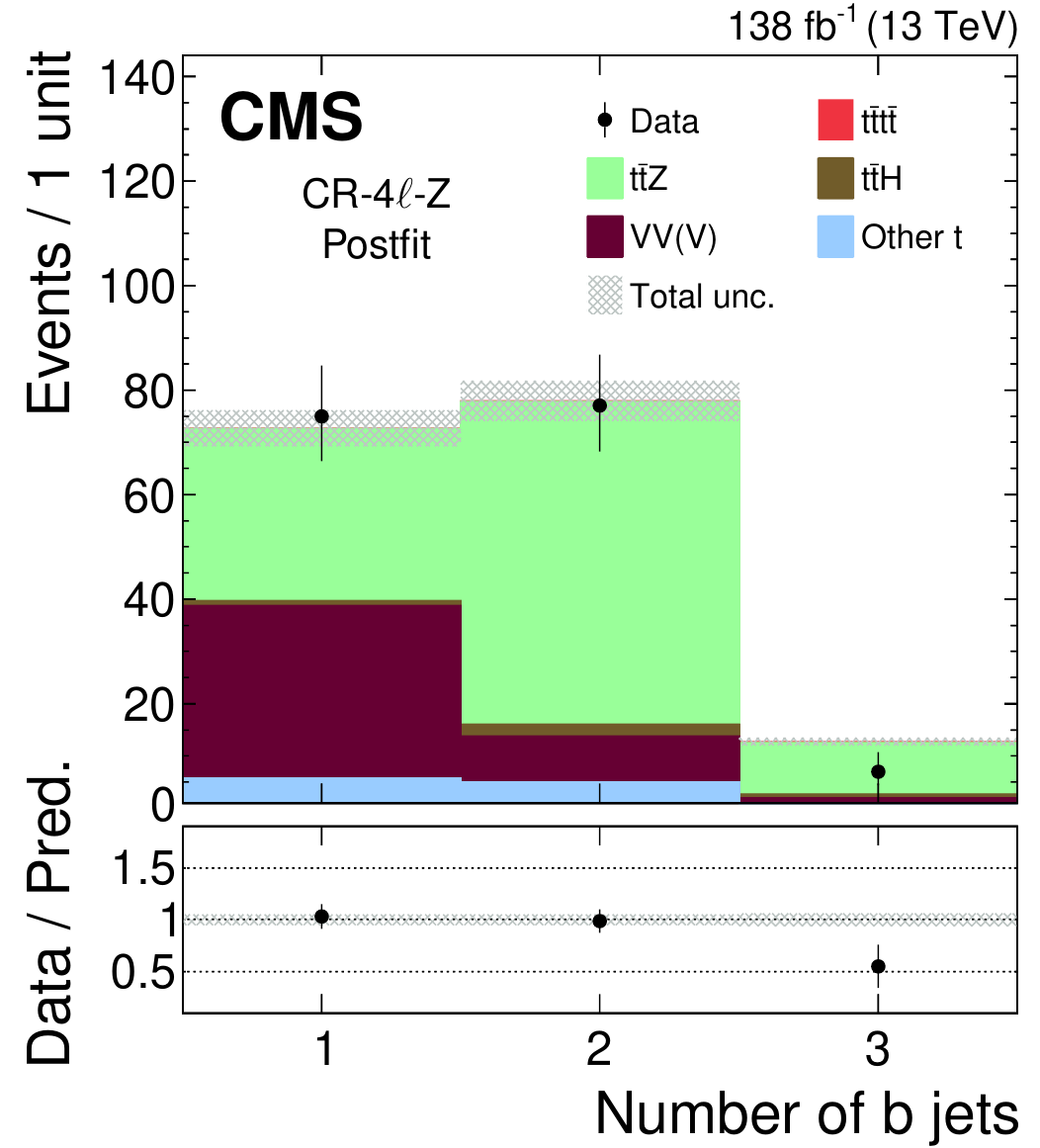}%
\caption{%
    Comparison of the number of observed (points) and predicted (colored histograms) events in the number of jets distribution in \CRthreeZ (left), and in the number of \PQb jets distribution in \CRthreeZ (middle) and \CRfourZ (right).
    The vertical bars on the points represent the statistical uncertainties in the data, and the hatched bands the total uncertainty in the predictions.
    The signal and background yields are shown with their best fit normalizations from the simultaneous fit to the data (``postfit'').
}
\label{fig:CR-Z}
\end{figure*}

Nonprompt-background contributions are estimated with a ``tight-to-loose'' ratio method~\cite{CMS:SUS-15-008, CMS:HIG-19-008, CMS:TOP-21-011}.
The probabilities for a loose lepton to also pass the tight ID are measured, separately for electrons and muons, as functions of \pt and \abseta in a data sample enriched in events composed uniquely of jets produced through the strong interaction, such that most selected leptons in this sample are nonprompt.
The measured probability is then applied to data events in a sideband of the SRs and CRs where one or more of the selected leptons fail the tight ID while passing the loose ID.
The method is validated in simulation separately with \ttbar and \Zjets production events, which have different main sources of nonprompt leptons, and also separately for nonprompt electrons and muons.
We compare the event yields obtained from simulation with two methods:\ either directly using nonprompt leptons that pass the tight ID, or using events with nonprompt leptons that pass the loose but fail the tight ID and applying the weights from the tight-to-loose ratio method.
Good agreement of better than 30\% is found in the most relevant kinematic distributions.
A normalization uncertainty of 30\% is assigned to the nonprompt background to account for any possible bias in the method.
Additionally, we compare data and predicted background yields in \CRtwoLow, \CRtwoHigh, and \CRthreeLow, which have relevant nonprompt-background contributions, and find good agreement within the assigned uncertainty.
The agreement in three different selections enriched in nonprompt-lepton backgrounds is shown in Fig.~\ref{fig:CR-Nonprompt}.

\begin{figure*}[!tb]
\centering
\includegraphics[width=0.32\textwidth]{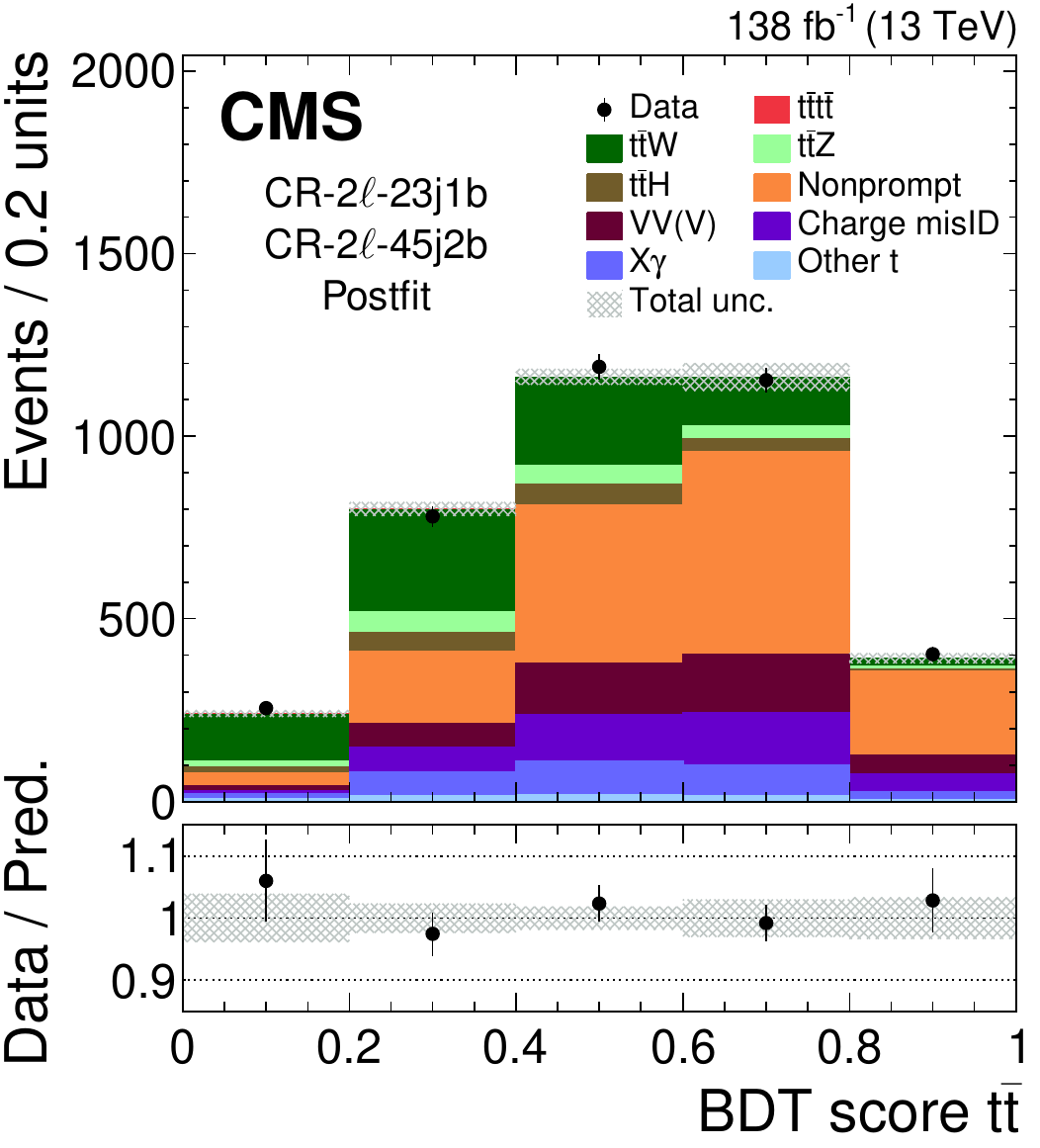}%
\hfill%
\includegraphics[width=0.32\textwidth]{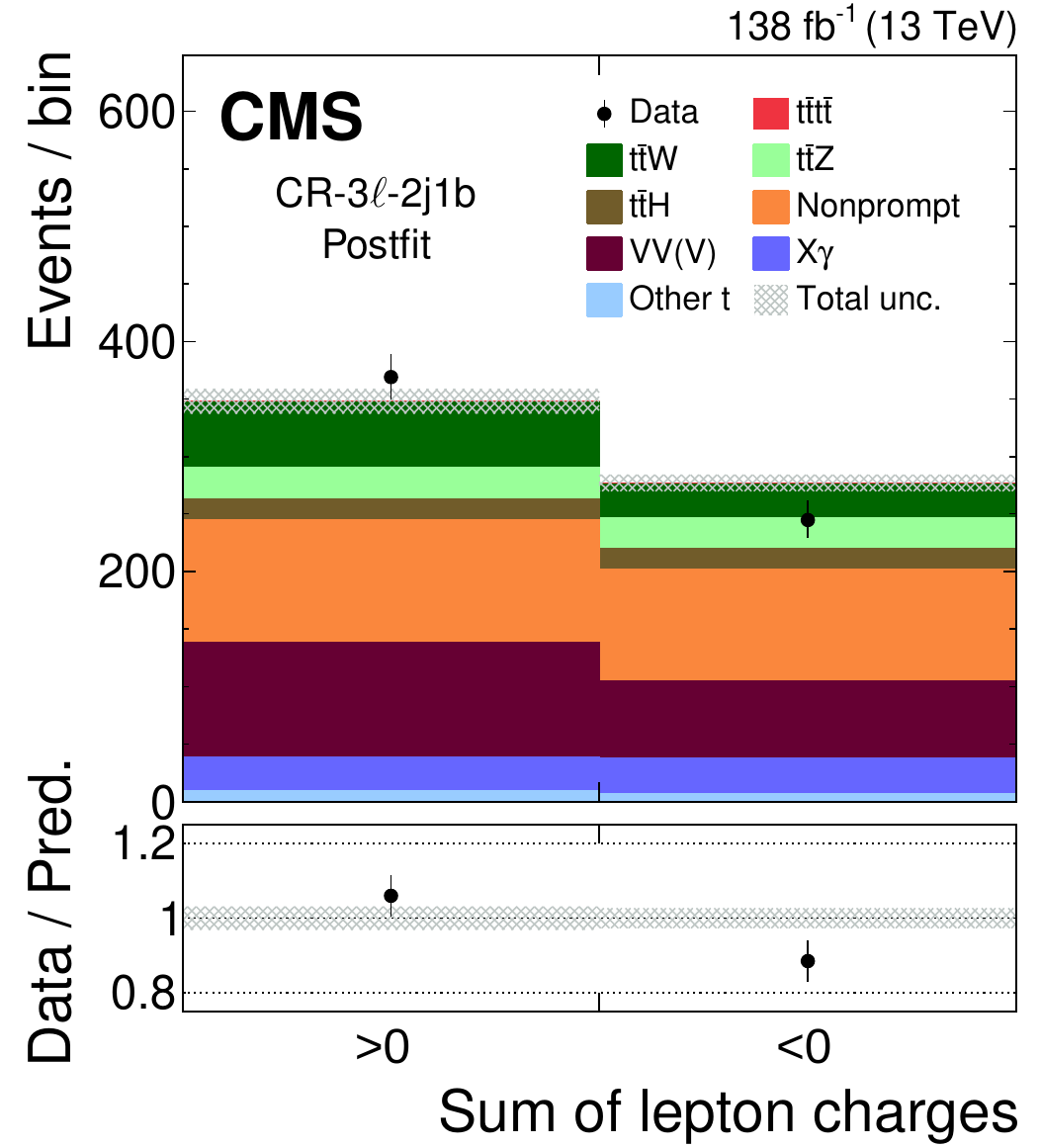}%
\hfill%
\includegraphics[width=0.32\textwidth]{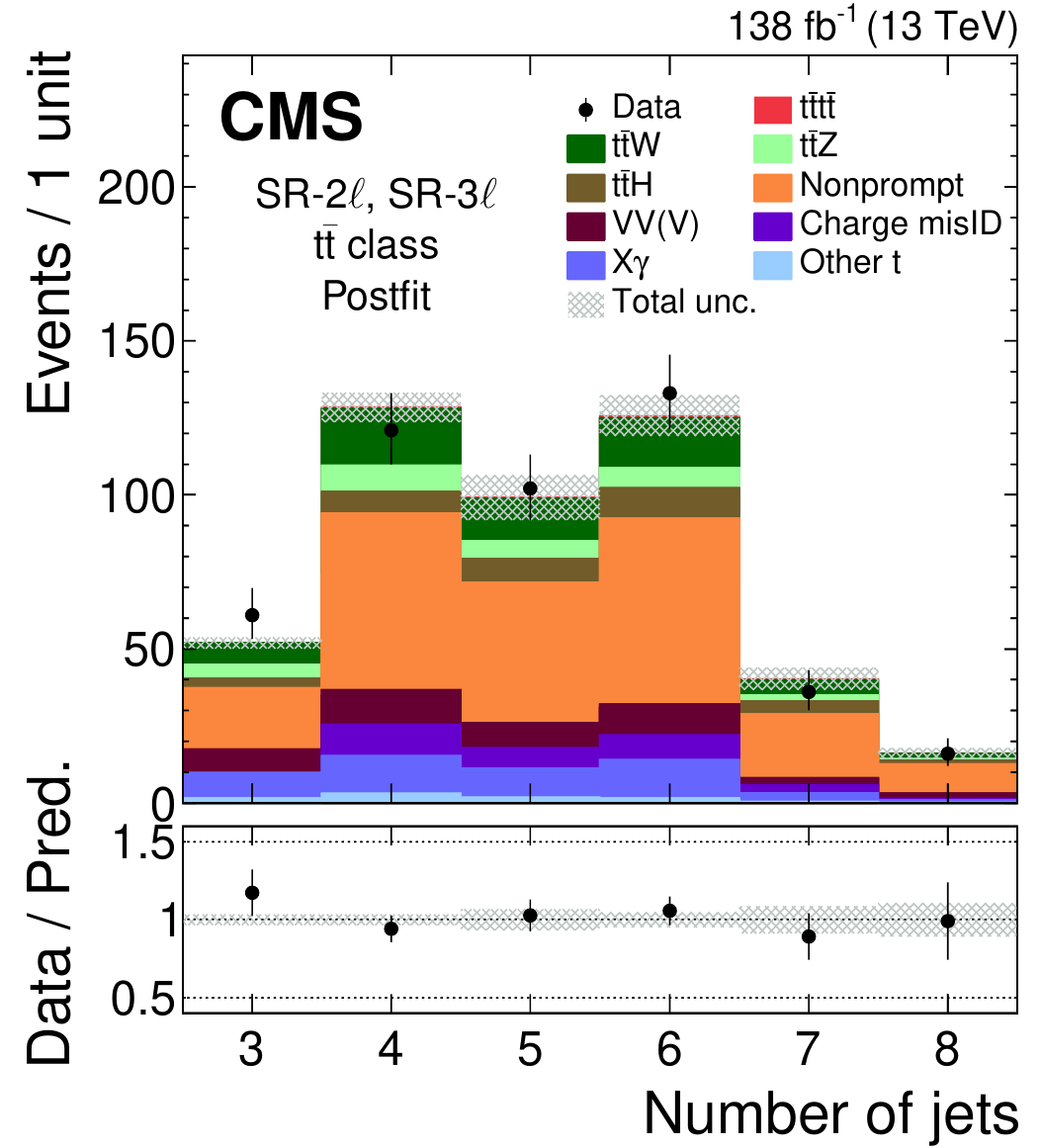}%
\caption{%
    Comparison of the number of observed (points) and predicted (colored histograms) events in the BDT score \ttbar in the combined \CRtwoLow and \CRtwoHigh (left), in the event yields with positive and negative sum of lepton charges in \CRthreeLow (middle), and in the number of jets distribution in the \ttbar class of the combined \SRtwo and \SRthree (right).
    The vertical bars on the points represent the statistical uncertainties in the data, and the hatched bands the total uncertainty in the predictions.
    The signal and background yields are shown with their best fit normalizations from the simultaneous fit to the data (``postfit'').
}
\label{fig:CR-Nonprompt}
\end{figure*}

To determine the charge-misID background, the electron charge-misID probability is determined in simulated \Zjets samples and parameterized as a function of \pt and \abseta.
This probability is applied as a scale factor to sideband events in data (\ie, with opposite-sign instead of same-sign leptons) to obtain a prediction of the charge-misID background in the \SRtwo.
A validation region (its sideband) is defined by requiring exactly two tight same-sign (opposite-sign) electrons that form a \PZ candidate.
To evaluate a possible mismodeling of the charge-misID probability in simulation, we evaluate the ratio of the sideband yields to the validation region yields both in simulation and data, and derive correction factors to the charge-misID probability from the data-to-simulation differences in this ratio.
The correction factors are evaluated separately for each year of data taking and integrated over electron \pt and \abseta.
After applying the scale and correction factors to the sideband event yields, the agreement with the validation region yields is better than 15\% across all relevant kinematic distributions.

\section{Systematic uncertainties}
\label{sec:systematics}

Several sources of systematic uncertainty affect the acceptance and reconstruction efficiency in simulation, the background event yields, and the distributions of the observables used for the signal extraction.
Experimental uncertainties in the integrated luminosity, lepton selection efficiency, trigger efficiency, JES, \PQb tagging efficiency, and nonprompt-background estimation are considered with several sources each, of which some are correlated and others uncorrelated between the years of data taking.
Only the experimental uncertainties from JER and unclustered \ptmiss contributions are fully uncorrelated between the data-taking years.
Uncertainties in the simulated pileup distribution, background normalization, and shapes of the fitted observables estimated from simulation are fully correlated between the data-taking years.

The measured integrated luminosity values for the three data-taking years have uncertainties between 1.2 and 2.5\%~\cite{CMS:LUM-17-003, CMS:LUM-17-004, CMS:LUM-18-002}, while the overall uncertainty for the combined data set is 1.6\%, affecting both the background predictions from simulation and the measurement of the \tttt cross section.
To evaluate the uncertainty in the distribution of pileup events in simulation, we vary the total \pp inelastic cross section by $\pm$4.6\%.

The efficiency of the tight ID selection of electrons and muons is measured in data and simulation with the ``tag-and-probe'' method in $\PZ\to\Pell^+\Pell^-$ events~\cite{CMS:EWK-10-002, CMS:EGM-17-001, CMS:MUO-16-001}.
Per-lepton correction factors are derived and applied to simulated events, and statistical and systematic uncertainties are considered, with overall uncertainty values of typically below 2.6 (0.5)\% per electron (muon) with $\pt>20\GeV$, and 3.5 (1.2)\% with $10<\pt<20\GeV$.
We have validated the correction factors by comparing data and simulated \ttbar samples in $\Pepm\PGmmp$ events with at least one medium \PQb jet, observing good agreement within the assigned uncertainties and thus confirming that the correction factors are applicable for a phase space with a larger jet multiplicity.
The trigger efficiency is measured by selecting events with two or three leptons in an unbiased data sample, collected with triggers on \ptmiss or hadronic activity, and corrections are derived to match the efficiency in simulation to that observed in data.
A systematic uncertainty of 3\% is assigned to simulated event yields to cover residual differences, and additionally the statistical uncertainties in the measurement are considered.
Special correction factors are applied for a gradual shift in the timing of the ECAL inputs to the first-level trigger, which caused a specific trigger inefficiency during the 2016 and 2017 data-taking periods~\cite{CMS:TRG-17-001}.

The JES and JER uncertainties are taken into account by applying variations to the energy of the reconstructed jets in simulated events, considering various uncertainty sources split between detector regions and data-taking years~\cite{CMS:JME-13-004}.
The JES variations, as well as an additional variation to account for uncertainty contributions from unclustered PF particles~\cite{CMS:JME-17-001}, are also propagated to the \ptmiss.
Corrections are applied to simulated events to match the shape of the \DeepJet discriminant in simulation to that in data, and uncertainties in the correction factors are evaluated separately for heavy-flavor and light jets~\cite{CMS:BTV-16-002}.

Uncertainties in the background predictions include normalization uncertainties of 7.7\% for \ttH~\cite{Kulesza:2020nfh}, 6.2\% for \WZ~\cite{CMS:SMP-18-002}, 20\% for \ttt~\cite{Durieux:2023ttt}, 20\% for \otherT~\cite{CMS:TOP-20-010}, 6\% for \VVV~\cite{CMS:SMP-16-017, CMS:SMP-18-004}, and 5\% for \Xgamma~\cite{CMS:TOP-21-004}.
The normalization of the \ttZ and \ttW background contributions are not constrained a priori but treated as free parameters in the fit.
For the \WZ and \ZZ backgrounds, additional normalization uncertainties corresponding to the statistical uncertainties in the scale factors per \Njets-bin are assigned.
For the nonprompt-background prediction, the statistical uncertainties in the measured tight-to-loose ratio and the sideband region yields are considered, as well as an overall normalization uncertainty.
We assign normalization uncertainties separately for contributions with nonprompt electrons and muons.
To account both for possible systematic effects in the method itself and for potential year-to-year differences, we assign uncertainty components of 20\% each that are either fully correlated or uncorrelated between the data-taking years.
Adding these components in quadrature, the overall normalization uncertainty of 30\%, based on the validation tests discussed in Section~\ref{sec:background}, is retained.
For the charge-misID background, the statistical uncertainty in the event yields in the sideband regions and a normalization uncertainty of 15\% are considered.

Various modeling uncertainties are considered for the signal predictions as well as for all simulated background predictions, including the \ttZ and \ttW background contributions whose normalization is determined by free parameters in the fit.
The choice of the renormalization and factorization scales in the matrix-element calculations are assessed by individual and simultaneous variations up and down by a factor of 2 (excluding the two extreme variations)~\cite{EXT:TtbarSystematic-2004}, and evaluating the envelope of the six obtained variations in the fitted distributions.
This is done separately for each process.
The limited knowledge of the proton PDFs is taken into account using NNPDF replicas~\cite{EXT:NNPDF-2017} and applying the procedure described in Ref.~\cite{Butterworth:2015oua}, resulting in 100 individual variations that are evaluated simultaneously for all processes.
For the choice of the renormalization scale for QCD emissions in initial-state (final-state) radiation in the parton shower simulation, individual variations by a factor of 2 up and down are considered, separately (simultaneously) for the considered processes.
The variations associated with the modeling uncertainties generally affect the overall normalization, the simulated acceptance, and the shapes of the fitted distributions.
Since background normalization uncertainties are assigned separately and no such uncertainty is required when measuring the cross section of the signal process, we remove the impact on the overall normalization of each process before any selection from these variations.

Additional uncertainties are considered to account for a possible mismodeling in the jet multiplicity in \ttX production processes.
Motivated by dedicated studies of the cross section ratio of \ttbar production in association with additional \PQb and light jets~\cite{CMS:TOP-18-002} that find an underprediction of \ttbb events in simulated \ttbar event samples, we assign a conservative uncertainty of 40\% as additional normalization uncertainty to the contributions of \ttW, \ttZ, and \ttH production with at least one additional particle-level jet containing a \PQb hadron, identified using a ghost-matching procedure~\cite{CMS:NOTE-2017-004}, with a distance of $\DR>0.4$ to any parton-level \PQb quark from a top quark decay.
This uncertainty is treated as uncorrelated between these processes.
Dedicated theoretical studies have been performed to evaluate the possible mismodeling in the number of additional jets in \ttW production~\cite{Frederix:2021agh}.
We include a one-sided variation of the \Njets distributions in \ttW production, derived as the difference between the normalized distributions predicted by the simulation used in this analysis and by Ref.~\cite{Frederix:2021agh}, as additional uncertainty.
The scale factors describing this variation are between 0.92 at $\Njets=2$ and 1.55 at $\Njets\geq7$ in the \twol channel, and between 0.96 at $\Njets=2$ and 1.44 at $\Njets\geq6$ in the \threel channel.

\section{Results}

The cross sections of the \tttt signal process and the \ttW and \ttZ background processes are extracted from a simultaneous binned profile likelihood fit to the data in the SRs and CRs~\cite{CMS:NOTE-2011-005, EXT:Asymptotic-2011}.
A binned likelihood function $\likeli(\sigstr,\nuisan)$ is constructed as the product of the Poisson probabilities to obtain the observed yields given the predicted signal and background estimates in each bin, and includes all sources of systematic uncertainty as nuisance parameters denoted as \nuisan.
The set \sigstr denotes the signal strength modifiers that scale the normalizations of the predicted \tttt, \ttW, and \ttZ estimates, and the measured cross sections are obtained by multiplying the predicted cross sections of the three processes with the fitted value of the respective modifier.
Statistical uncertainties in the predicted yields are implemented through a single nuisance parameter in each bin for all processes~\cite{Barlow:1993dm, Conway:2011in}.
The fitted distributions are the BDT scores in the SR classes, the BDT score \ttbar in \CRtwoLow and \CRtwoHigh, the \Njets distribution in \CRthreeZ and \CRfourZ, and the event yields with positive or negative sum of lepton charges in \CRthreeLow.
In \SRtwo, the \tttt and \ttX classes are also split according to the lepton flavors to achieve a better separation of the nonprompt, charge-misID, and \Xgamma backgrounds.
Different choices of fit distributions and splits were evaluated, and this selection was found to give the highest expected significance for the \tttt signal.
Separate distributions per data-taking year are used in the fit.
In Fig.~\ref{fig:SR-tttt}, the yields in the SR \tttt classes after the fit are shown for all three years combined, using the same binning as for the fit.
A comparison of the yields in the \SRtwo and \SRthree \tttt classes before and after the fit is given in Table~\ref{tab:yields}.
No events are observed in data in the \SRfour \tttt class.

\begin{figure*}[!t]
\centering
\includegraphics[width=0.32\textwidth]{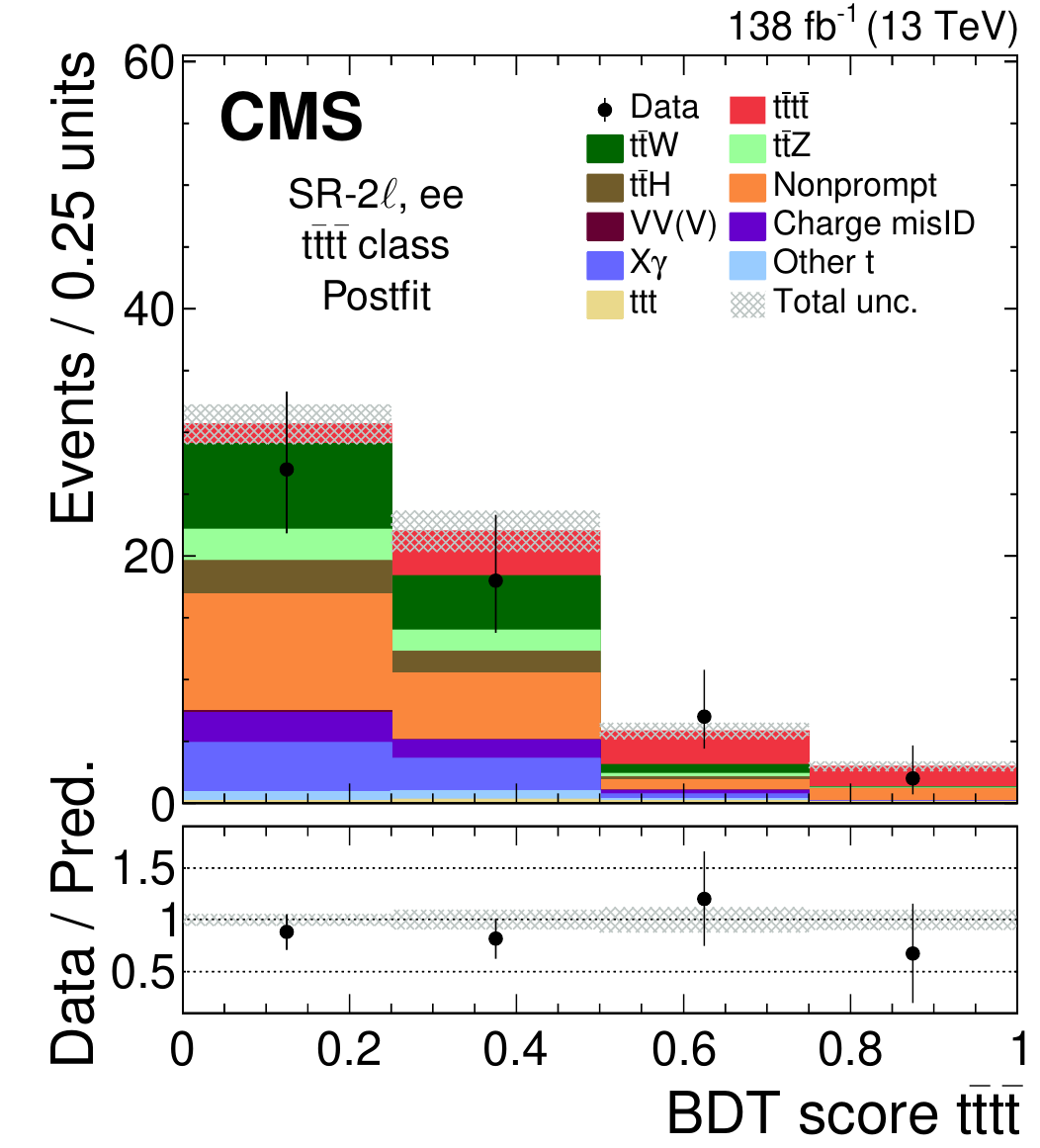}%
\hfill%
\includegraphics[width=0.32\textwidth]{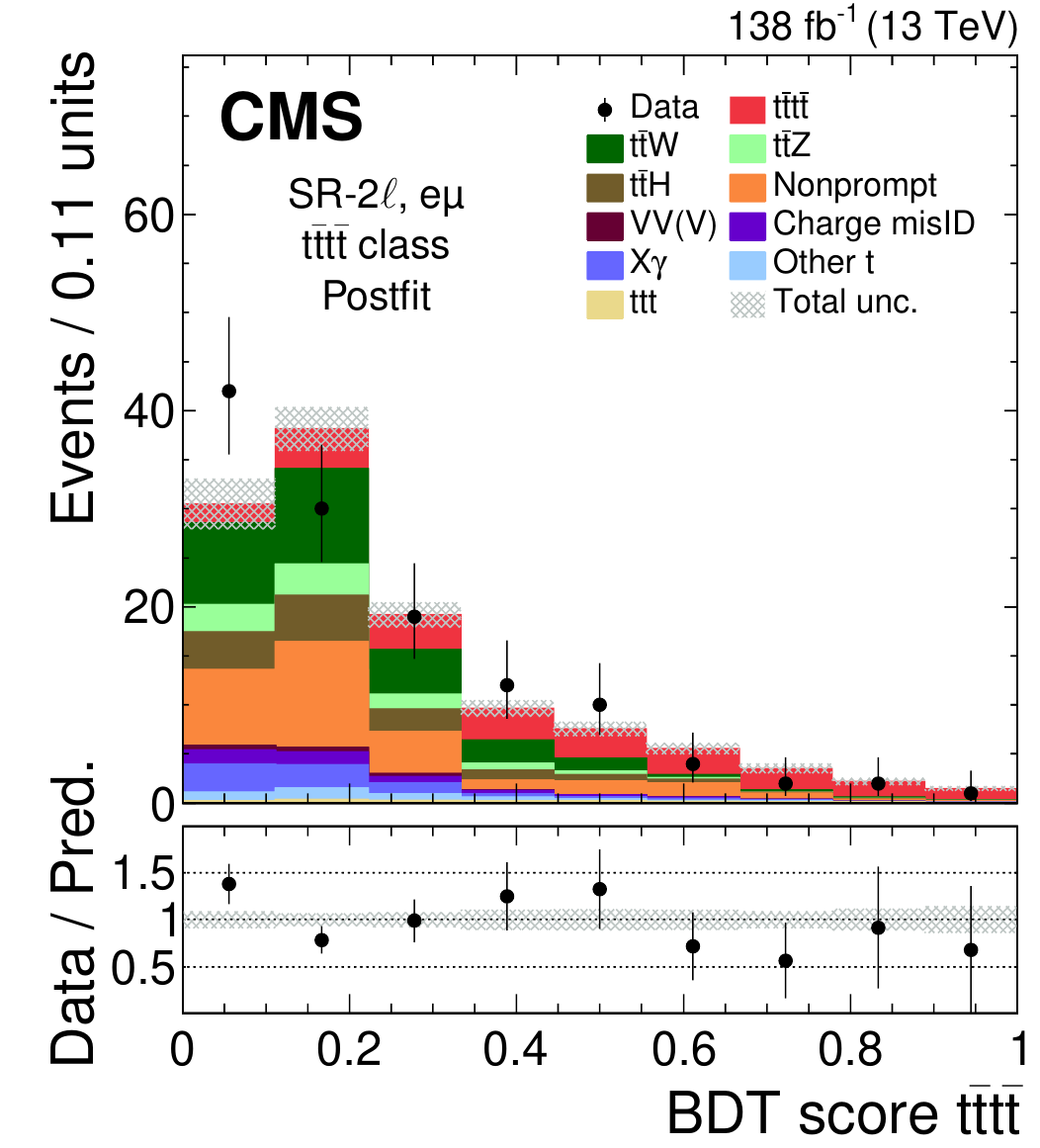}%
\hfill%
\includegraphics[width=0.32\textwidth]{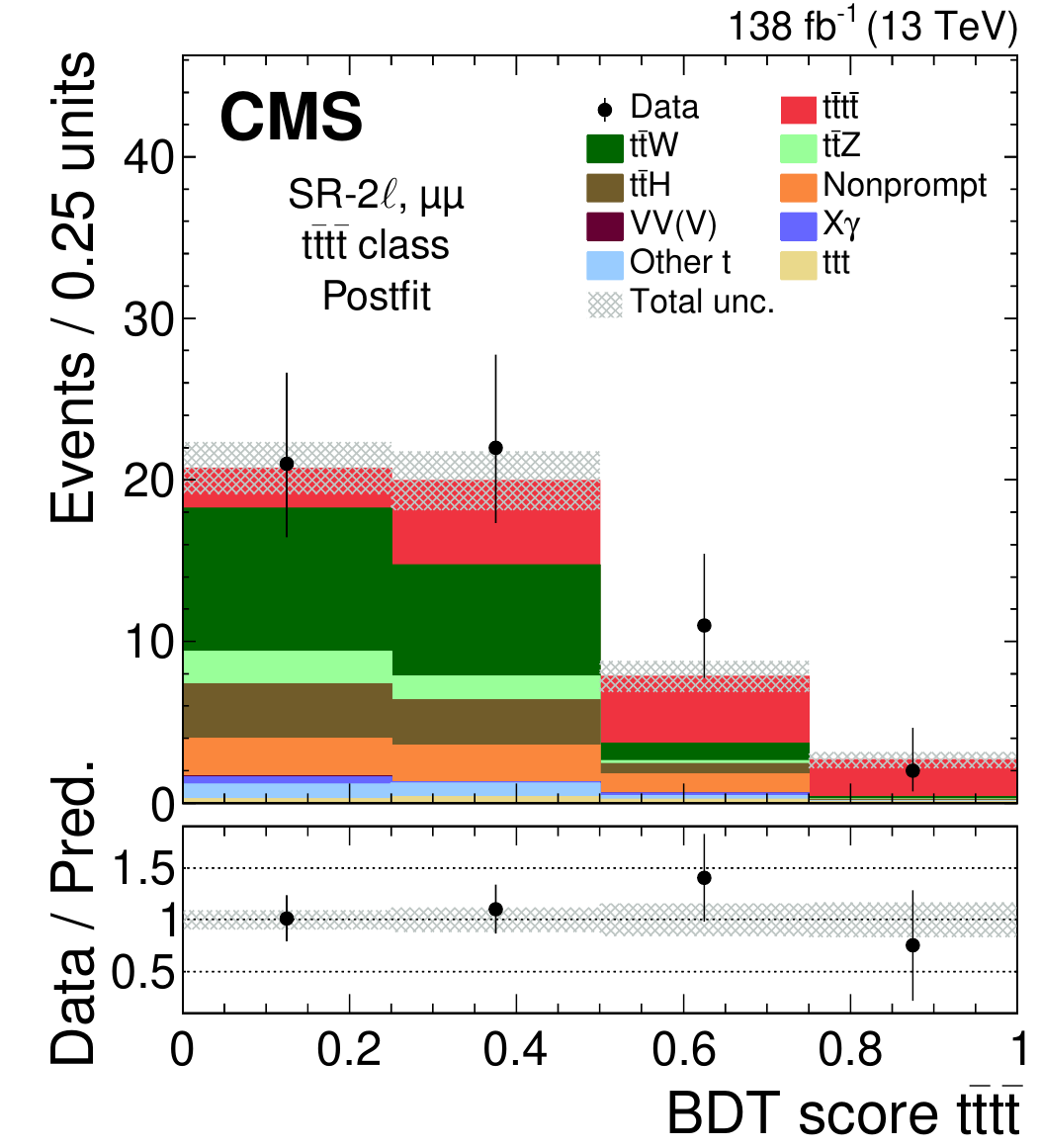} \\
\includegraphics[width=0.32\textwidth]{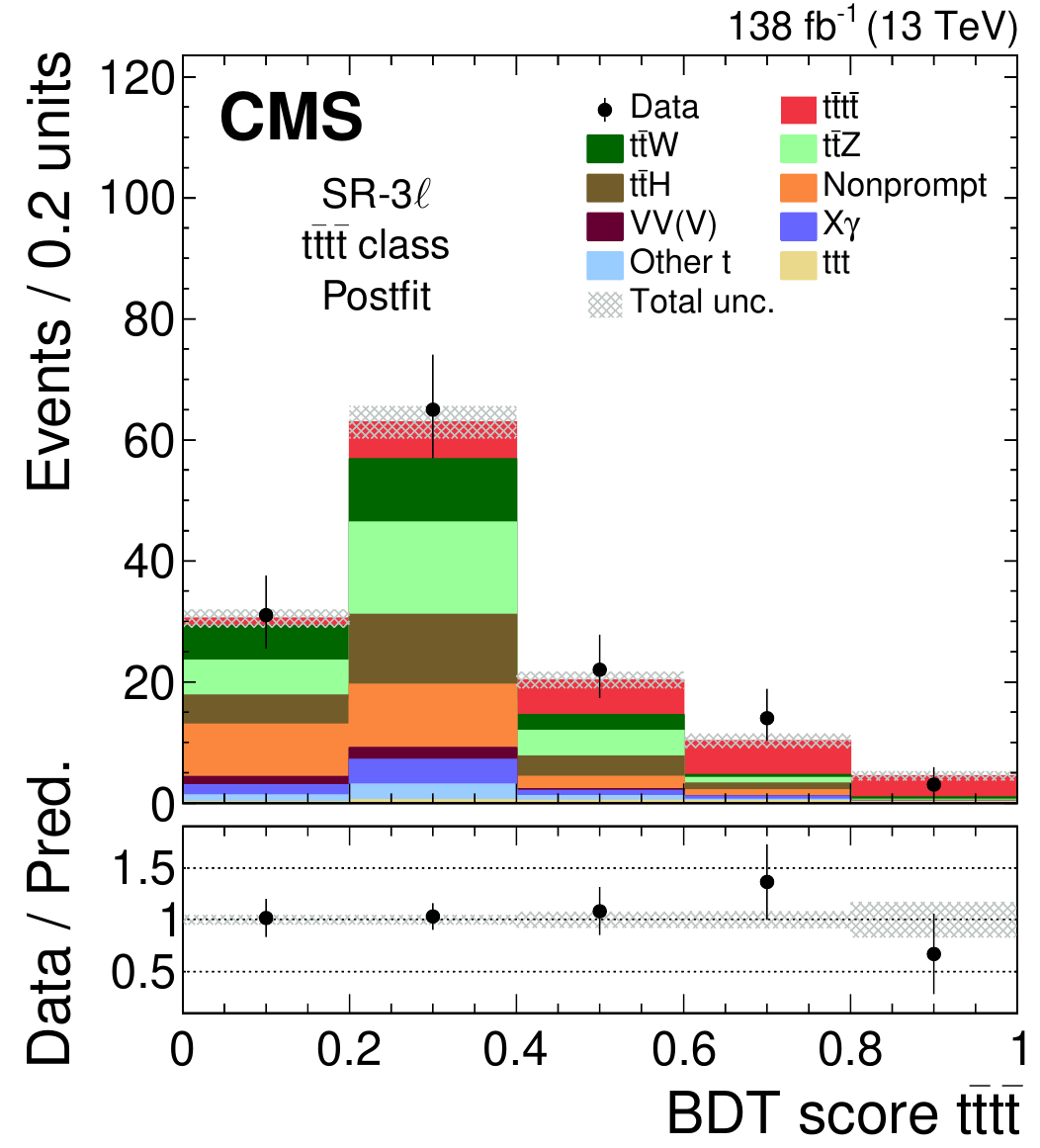}%
\hfill%
\includegraphics[width=0.32\textwidth]{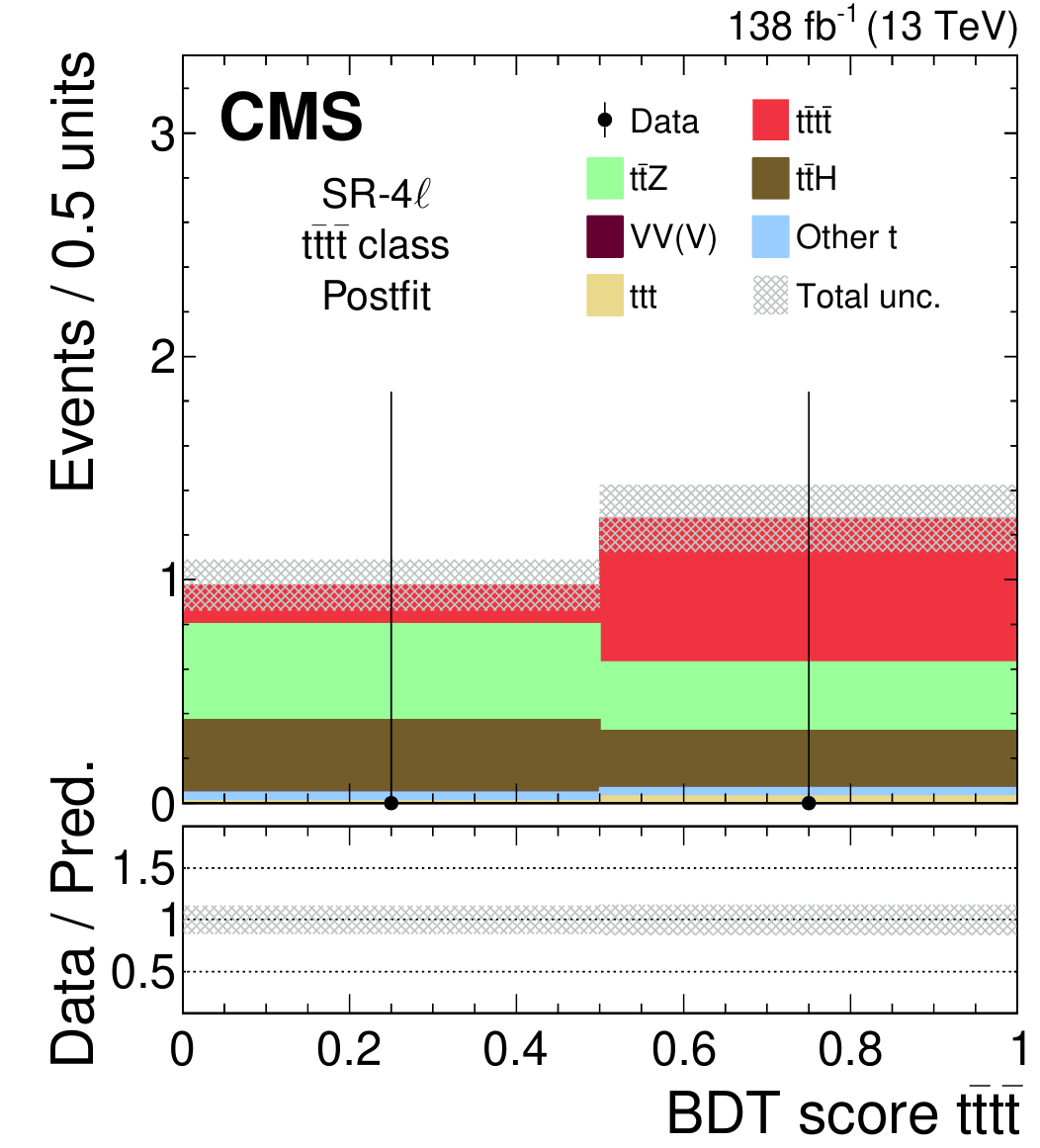}%
\hfill%
\includegraphics[width=0.32\textwidth]{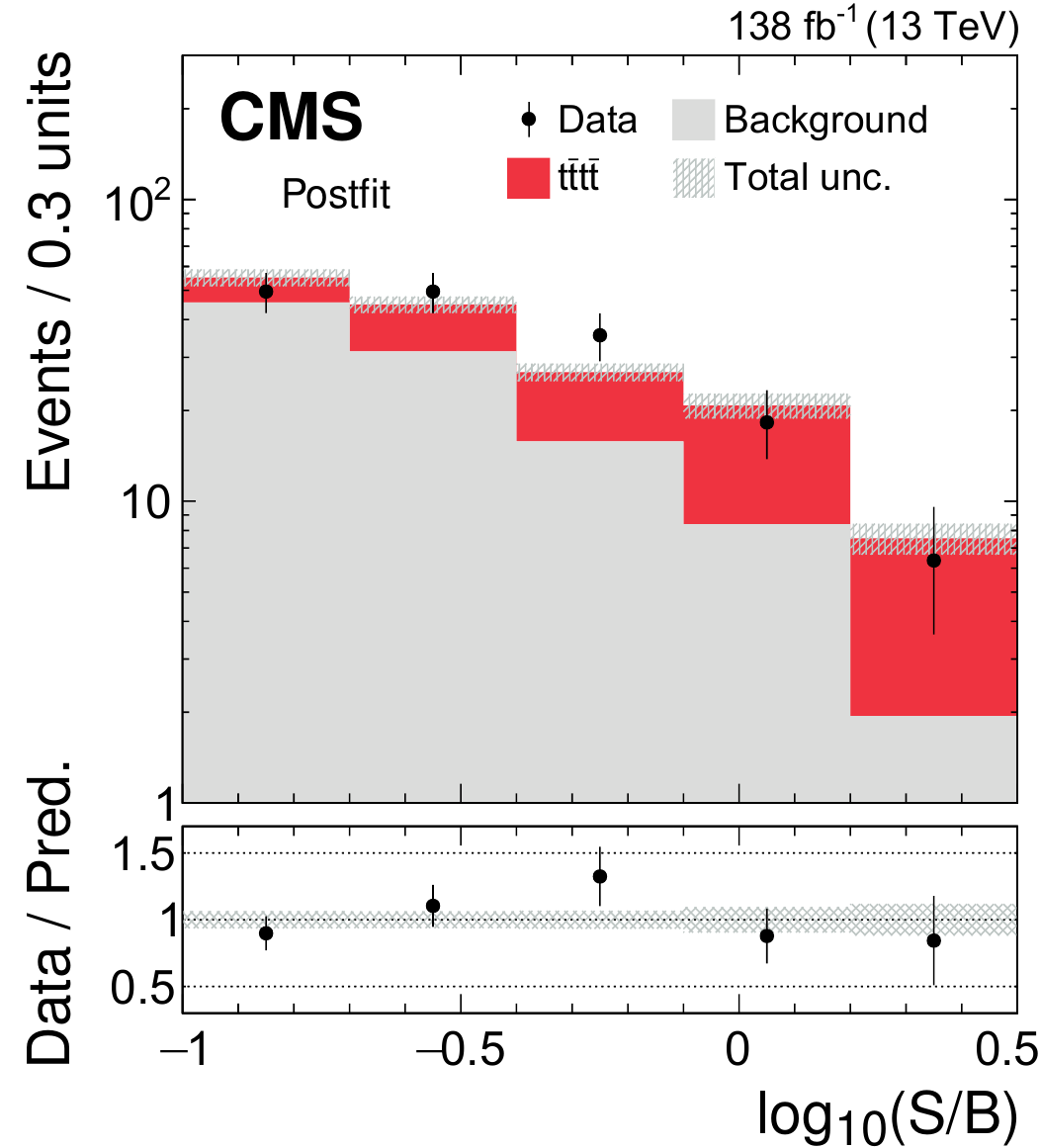}%
\caption{%
    Comparison of the number of observed (points) and predicted (colored histograms) events in the BDT score \tttt in the \tttt classes of \SRtwo, shown for the $\Pe\Pe$ (upper left), $\Pe\PGm$ (upper middle), and \mumu (upper right) categories, of \SRthree (lower left) and of \SRfour (lower middle).
    Additionally, the comparison is shown for all SRs combined as a function of \logsoverb (lower right), where S and B are evaluated for each bin of the fitted distributions as the predicted signal and background yields before the fit to data.
    Bins with $\logsoverb<-1$ are not included, and bins with $\logsoverb>0.5$ are included in the last bin.
    The vertical bars on the points represent the statistical uncertainties in the data, and the hatched bands the total uncertainty in the predictions.
    The signal and background yields are shown with their best fit normalizations from the simultaneous fit to the data (``postfit'').
}
\label{fig:SR-tttt}
\end{figure*}

The test statistic $\testst=-2\ln\likeli(\sigstr,\nuisanHatR)/\likeli(\sigstrHat,\nuisanHat)$ is constructed from the sets \sigstrHat and \nuisanHat that simultaneously maximize \likeli, and using the set \nuisanHatR that maximizes \likeli for a fixed set \sigstr.
We employ a calculation that simplifies \testst using an asymptotic approximation~\cite{CMS:NOTE-2011-005, EXT:Asymptotic-2011}, or a full calculation of \testst with toys for the fit using only the \fourl channel.
The observed (expected) statistical significance of the \tttt signal is found to be \sigobs (\sigexp) SDs from the background-only hypothesis.
The results when fitting each channel on its own are summarized in Fig.~\ref{fig:perchannel}, with the largest sensitivity provided by the \twol channel.
This constitutes an observation of \tttt production.
The cross section of \tttt production is measured to be
\begin{linenomath}\begin{equation*}
    \sigtttt=\xsecmeas=17.7\pmasym{+4.4}{-4.0}\fb.
\end{equation*}\end{linenomath}
This result is in agreement with the SM prediction of \xsecpred~\cite{vanBeekveld:2022hty} at the level of 1.0 SDs, when taking uncertainties of both prediction and measurement into account.

\begin{table*}[!p]
\centering
\topcaption{%
    Number of predicted and observed events in the \SRtwo and \SRthree \tttt classes, both before the fit to the data (``prefit'') and with their best fit normalizations (``postfit'').
    The uncertainties in the predicted number of events include both the statistical and systematic components.
    The uncertainties in the total number of predicted background and background plus signal events are also given.
    A dash indicates that the corresponding background does not contribute.
}
\label{tab:yields}
\renewcommand{\arraystretch}{1.2}
\begin{tabular}{c*{4}{r@{\,$\pm$\,}l}}
    \multirow{2}{*}{Process} & \multicolumn{4}{c}{\SRtwo \tttt class} & \multicolumn{4}{c}{\SRthree \tttt class} \\[-2pt]
    & \multicolumn{2}{c}{Prefit} & \multicolumn{2}{c}{Postfit} & \multicolumn{2}{c}{Prefit} & \multicolumn{2}{c}{Postfit} \\
    \hline
    \tttt & 35.85 & 0.85 & 47.4 & 6.4 & 16.92 & 0.38 & 22.4 & 3.1 \\ [0.3\cmsTabSkip]
    \ttt & 3.05 & 0.36 & 3.06 & 0.43 & 1.44 & 0.17 & 1.45 & 0.21 \\
    \ttW & 40.8 & 6.6 & 55.9 & 6.9 & 14.5 & 1.9 & 19.5 & 2.1 \\
    \ttZ & 14.3 & 1.3 & 17.2 & 1.2 & 22.6 & 1.5 & 26.3 & 1.4 \\
    \ttH & 24.3 & 2.4 & 24.9 & 2.4 & 21.0 & 1.9 & 21.0 & 1.6 \\
    Other \PQt & 7.06 & 0.90 & 7.23 & 0.96 & 4.70 & 0.57 & 4.81 & 0.58 \\
    \WZ & 1.21 & 0.66 & 1.77 & 0.71 & 2.61 & 0.92 & 3.3 & 1.1 \\
    \VVV & 0.38 & 0.11 & 0.38 & 0.08 & 0.26 & 0.11 & 0.32 & 0.16 \\
    \Xgamma & 15.0 & 1.7 & 15.0 & 1.6 & 6.87 & 0.64 & 7.11 & 0.63 \\
    Charge misID & 8.24 & 0.68 & 8.29 & 0.63 & \multicolumn{2}{c}{\NA} & \multicolumn{2}{c}{\NA} \\
    Nonprompt \Pe & 32.0 & 5.0 & 37.7 & 3.2 & 13.3 & 2.5 & 16.4 & 1.5 \\
    Nonprompt \PGm & 10.2 & 2.2 & 12.0 & 1.5 & 4.1 & 1.2 & 5.00 & 0.70 \\ [0.3\cmsTabSkip]
    Total background & 157 & 17 & 183 & 15 & 91.2 & 7.7 & 105.2 & 4.6 \\
    Total prediction & 192 & 17 & 231 & 15 & 108.2 & 8.0 & 127.6 & 5.5 \\ [0.3\cmsTabSkip]
    Data & \multicolumn{4}{c}{232} & \multicolumn{4}{c}{135} \\
\end{tabular}
\end{table*}

\begin{figure*}[!p]
\centering
\includegraphics[width=0.67\textwidth]{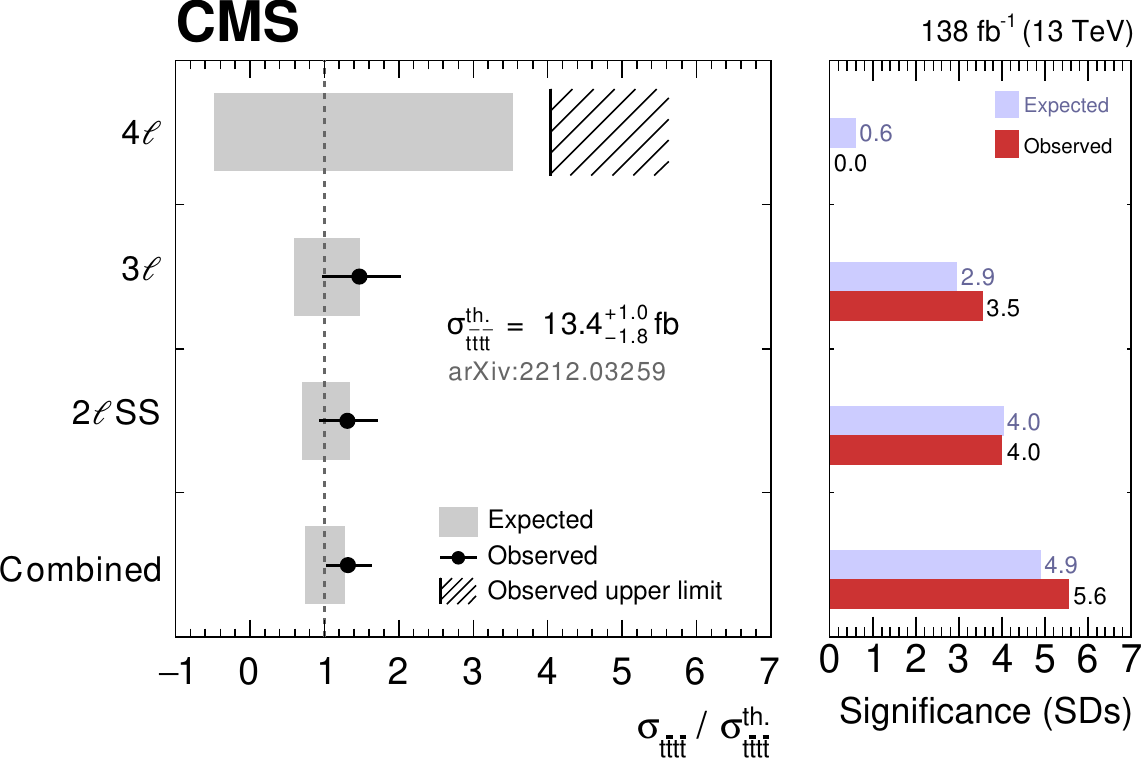}
\caption{%
    Comparison of fit results in the channels individually and in their combination.
    The left panel shows the values of the measured cross section relative to the SM prediction from Ref.~\cite{vanBeekveld:2022hty}, where the displayed uncertainty does not include the uncertainty in the SM prediction.
    For the \fourl channel, no events were observed in the \tttt-enriched bins, and an upper limit at 95\% confidence level is quoted.
    The right panel shows the expected and observed significance, with the printed values rounded to the first decimal.
}
\label{fig:perchannel}
\end{figure*}

The measurement of \sigtttt is limited by the statistical uncertainty in the SR event yields.
The largest systematic uncertainties arise from the \DeepJet correction factors and JES.
The largest uncertainty sources related to background modeling and simulation are the additional jets and \PQb jets in \ttW production, the impact of the \ttZ normalization, and the modeling of the \tttt signal process.
The values of the nuisance parameters associated with additional jets or \PQb jets in \ttX production after the fit are close to zero.
The 20 leading nuisance parameters in the fit are shown in Fig.~\ref{fig:impacts}.

\begin{figure*}[!ht]
\centering
\includegraphics[width=0.9\textwidth]{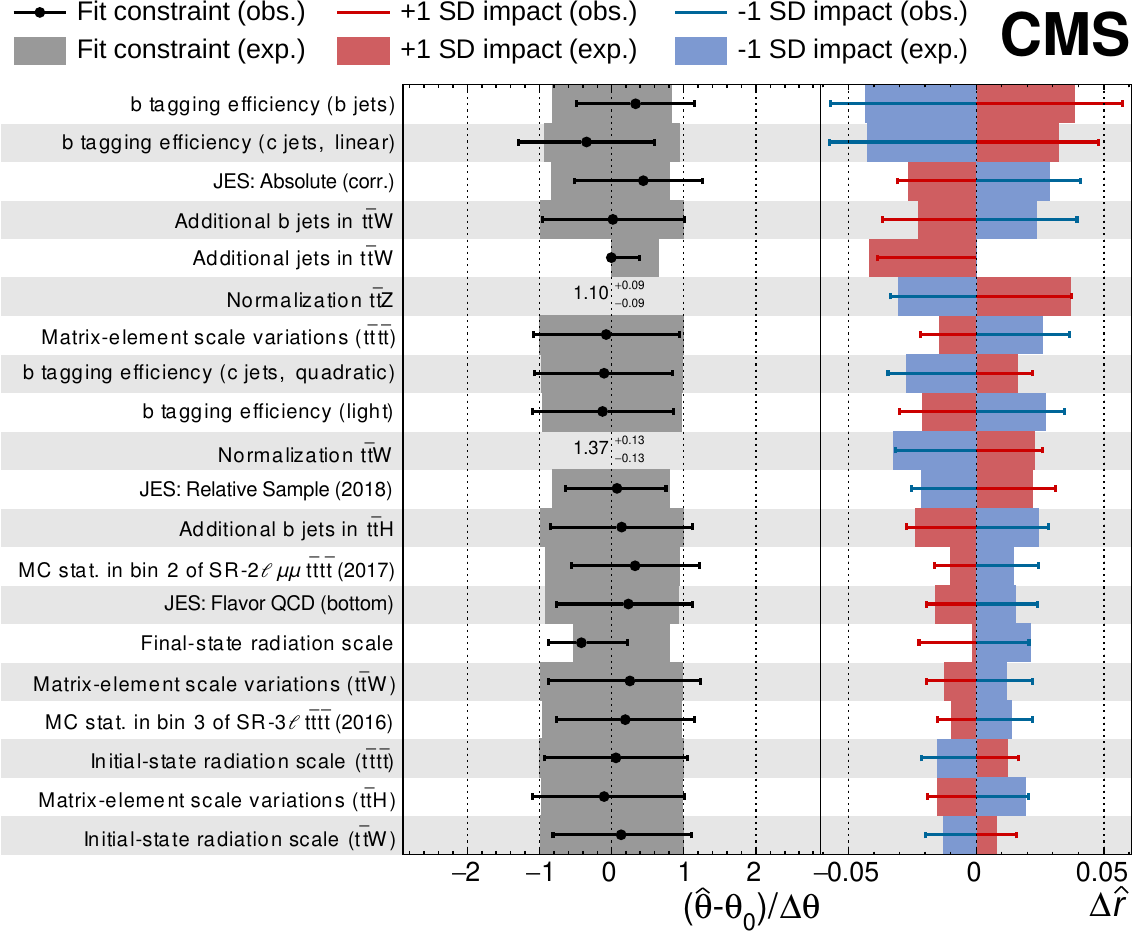}
\caption{%
    For the nuisance parameters listed in the left column, the pulls \nuisanPull (middle column) and impacts \sigstrHatDel (right column) are displayed.
    The 20 nuisance parameters with the largest impacts in the fit used to determine the \tttt cross section are shown.
    The impact \sigstrHatDel is obtained from varying the nuisance parameter \nuisan by $\pm1$ SD and evaluating the induced shift in the \tttt signal strength \sigstr.
    The pull \nuisanPull is calculated from the values \nuisanHat and \nuisanZero after and before the fit of \nuisan, respectively, and from its uncertainty \nuisanDel before the fit.
    The label ``corr.'' and the per-year labels indicate nuisance parameters associated with the correlated and uncorrelated parts of a systematic uncertainty, respectively.
    The nuisance parameters labeled ``MC stat.''\ refer to the per-bin statistical uncertainties in the predicted yields.
    The uncertainty associated with additional jets in \ttW production corresponds to a one-sided variation of the nominal template before the fit, and thus a one-sided impact after the fit is expected.
}
\label{fig:impacts}
\end{figure*}

The normalizations of the \ttW and \ttZ backgrounds are determined as free parameters in the fit and, when extrapolated to the full phase space, result in the following cross sections:
\begin{linenomath}\begin{equation*}\begin{aligned}
    \sigma(\ttW)&=990\pm58\stat\pm79\syst\fb=990\pm98\fb, \\
    \sigma(\ttZ)&=945\pm43\stat\pm69\syst\fb=945\pm81\fb.
\end{aligned}\end{equation*}\end{linenomath}
The correlations of the three simultaneously measured cross sections are positive, with values of 7\% between \tttt and \ttW, 11\% between \tttt and \ttZ, and 12\% between \ttW and \ttZ.
The \ttW cross section result is larger than the SM prediction of \ttwpred~\cite{Frederix:2021agh} at the level of 2.2 SDs, and also larger by more than 1.0 SDs than the central value $868\pm40\stat\pm51\syst\fb$ from the dedicated CMS cross section measurement based on the same data set using same-sign \twol and \threel events~\cite{CMS:TOP-21-011}.
The measured \ttZ cross section is in agreement with the SM prediction of \ttzpred~\cite{Kulesza:2020nfh} at the level of 0.8 SDs, and also in good agreement with the dedicated CMS cross section measurement of $950\pm50\stat\pm60\syst\fb$ based on a partial data set of 77.5\fbinv using \threel and \fourl events~\cite{CMS:TOP-18-009}.

\section{Summary}

A measurement of the production of four top quarks (\tttt) in proton-proton collisions at \sqrts has been presented, using events with two same-sign, three, and four charged leptons (electrons and muons) and additional jets from a data set corresponding to an integrated luminosity of 138\fbinv recorded with the CMS detector at the LHC.
Multivariate discriminants are employed in the identification of prompt leptons and jets originating from the decay of \PQb hadrons, and to distinguish between selected events from the \tttt signal and the main background contributions.
A profile likelihood fit is performed to the data in signal and control regions for the extraction of the \tttt cross section.

The improvements in object identification and analysis strategy bring the sensitivity of the analysis to the observation level, with an observed (expected) significance of the \tttt signal above the background-only hypothesis of \sigobs (\sigexp) standard deviations.
The signal cross section is measured to be $\sigtttt=\xsecmeas$, in agreement with the available standard model predictions.
This result marks a significant milestone in the top quark physics program of the LHC.

\begin{acknowledgments}
We congratulate our colleagues in the CERN accelerator departments for the excellent performance of the LHC and thank the technical and administrative staffs at CERN and at other CMS institutes for their contributions to the success of the CMS effort. In addition, we gratefully acknowledge the computing centres and personnel of the Worldwide LHC Computing Grid and other centres for delivering so effectively the computing infrastructure essential to our analyses. Finally, we acknowledge the enduring support for the construction and operation of the LHC, the CMS detector, and the supporting computing infrastructure provided by the following funding agencies: BMBWF and FWF (Austria); FNRS and FWO (Belgium); CNPq, CAPES, FAPERJ, FAPERGS, and FAPESP (Brazil); MES and BNSF (Bulgaria); CERN; CAS, MoST, and NSFC (China); MINCIENCIAS (Colombia); MSES and CSF (Croatia); RIF (Cyprus); SENESCYT (Ecuador); MoER, ERC PUT and ERDF (Estonia); Academy of Finland, MEC, and HIP (Finland); CEA and CNRS/IN2P3 (France); BMBF, DFG, and HGF (Germany); GSRI (Greece); NKFIH (Hungary); DAE and DST (India); IPM (Iran); SFI (Ireland); INFN (Italy); MSIP and NRF (Republic of Korea); MES (Latvia); LAS (Lithuania); MOE and UM (Malaysia); BUAP, CINVESTAV, CONACYT, LNS, SEP, and UASLP-FAI (Mexico); MOS (Montenegro); MBIE (New Zealand); PAEC (Pakistan); MES and NSC (Poland); FCT (Portugal); MESTD (Serbia); MCIN/AEI and PCTI (Spain); MOSTR (Sri Lanka); Swiss Funding Agencies (Switzerland); MST (Taipei); MHESI and NSTDA (Thailand); TUBITAK and TENMAK (Turkey); NASU (Ukraine); STFC (United Kingdom); DOE and NSF (USA).

\hyphenation{Rachada-pisek} Individuals have received support from the Marie-Curie programme and the European Research Council and Horizon 2020 Grant, contract Nos.\ 675440, 724704, 752730, 758316, 765710, 824093, 884104, and COST Action CA16108 (European Union); the Leventis Foundation; the Alfred P.\ Sloan Foundation; the Alexander von Humboldt Foundation; the Belgian Federal Science Policy Office; the Fonds pour la Formation \`a la Recherche dans l'Industrie et dans l'Agriculture (FRIA-Belgium); the Agentschap voor Innovatie door Wetenschap en Technologie (IWT-Belgium); the F.R.S.-FNRS and FWO (Belgium) under the ``Excellence of Science -- EOS" -- be.h project n.\ 30820817; the Beijing Municipal Science \& Technology Commission, No. Z191100007219010; the Ministry of Education, Youth and Sports (MEYS) of the Czech Republic; the Hellenic Foundation for Research and Innovation (HFRI), Project Number 2288 (Greece); the Deutsche Forschungsgemeinschaft (DFG), under Germany's Excellence Strategy -- EXC 2121 ``Quantum Universe" -- 390833306, and under project number 400140256 - GRK2497; the Hungarian Academy of Sciences, the New National Excellence Program - \'UNKP, the NKFIH research grants K 124845, K 124850, K 128713, K 128786, K 129058, K 131991, K 133046, K 138136, K 143460, K 143477, 2020-2.2.1-ED-2021-00181, and TKP2021-NKTA-64 (Hungary); the Council of Science and Industrial Research, India; the Latvian Council of Science; the Ministry of Education and Science, project no. 2022/WK/14, and the National Science Center, contracts Opus 2021/41/B/ST2/01369 and 2021/43/B/ST2/01552 (Poland); the Funda\c{c}\~ao para a Ci\^encia e a Tecnologia, grant CEECIND/01334/2018 (Portugal); the National Priorities Research Program by Qatar National Research Fund; MCIN/AEI/10.13039/501100011033, ERDF ``a way of making Europe", and the Programa Estatal de Fomento de la Investigaci{\'o}n Cient{\'i}fica y T{\'e}cnica de Excelencia Mar\'{\i}a de Maeztu, grant MDM-2017-0765 and Programa Severo Ochoa del Principado de Asturias (Spain); the Chulalongkorn Academic into Its 2nd Century Project Advancement Project, and the National Science, Research and Innovation Fund via the Program Management Unit for Human Resources \& Institutional Development, Research and Innovation, grant B05F650021 (Thailand); the Kavli Foundation; the Nvidia Corporation; the SuperMicro Corporation; the Welch Foundation, contract C-1845; and the Weston Havens Foundation (USA).
\end{acknowledgments}

\bibliography{auto_generated}

\appendix
\numberwithin{figure}{section}
\section{Prefit distributions}
\ifthenelse{\boolean{cms@external}}{%
    \renewcommand{\thefigure}{\Alph{section}.\arabic{figure}}
    \setcounter{figure}{0}
}{}

See Figs.~\ref{fig:appendix1}--\ref{fig:appendix4}.

\begin{figure*}[!h]
\centering
\includegraphics[width=0.3\textwidth]{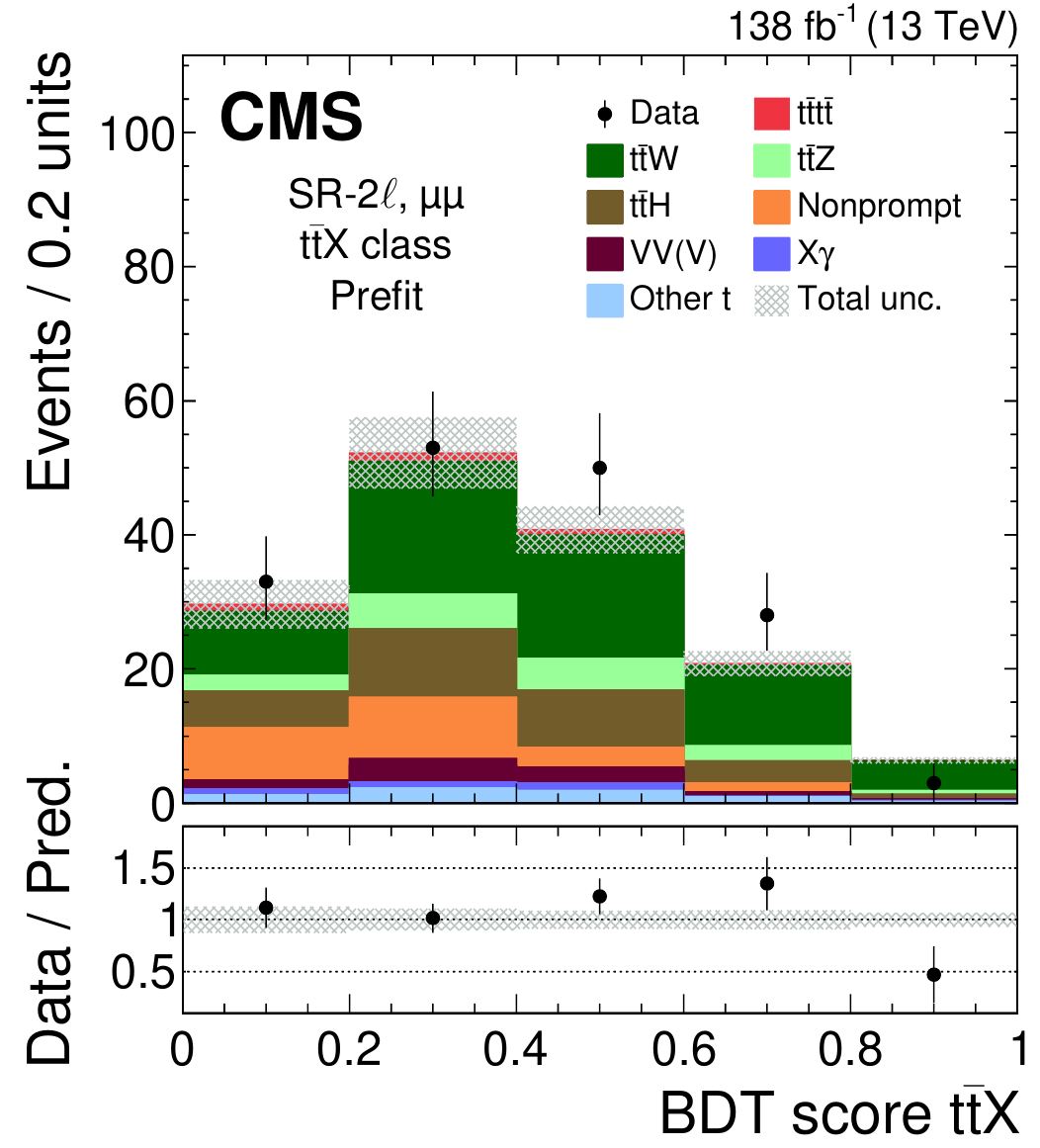}%
\hfill%
\includegraphics[width=0.3\textwidth]{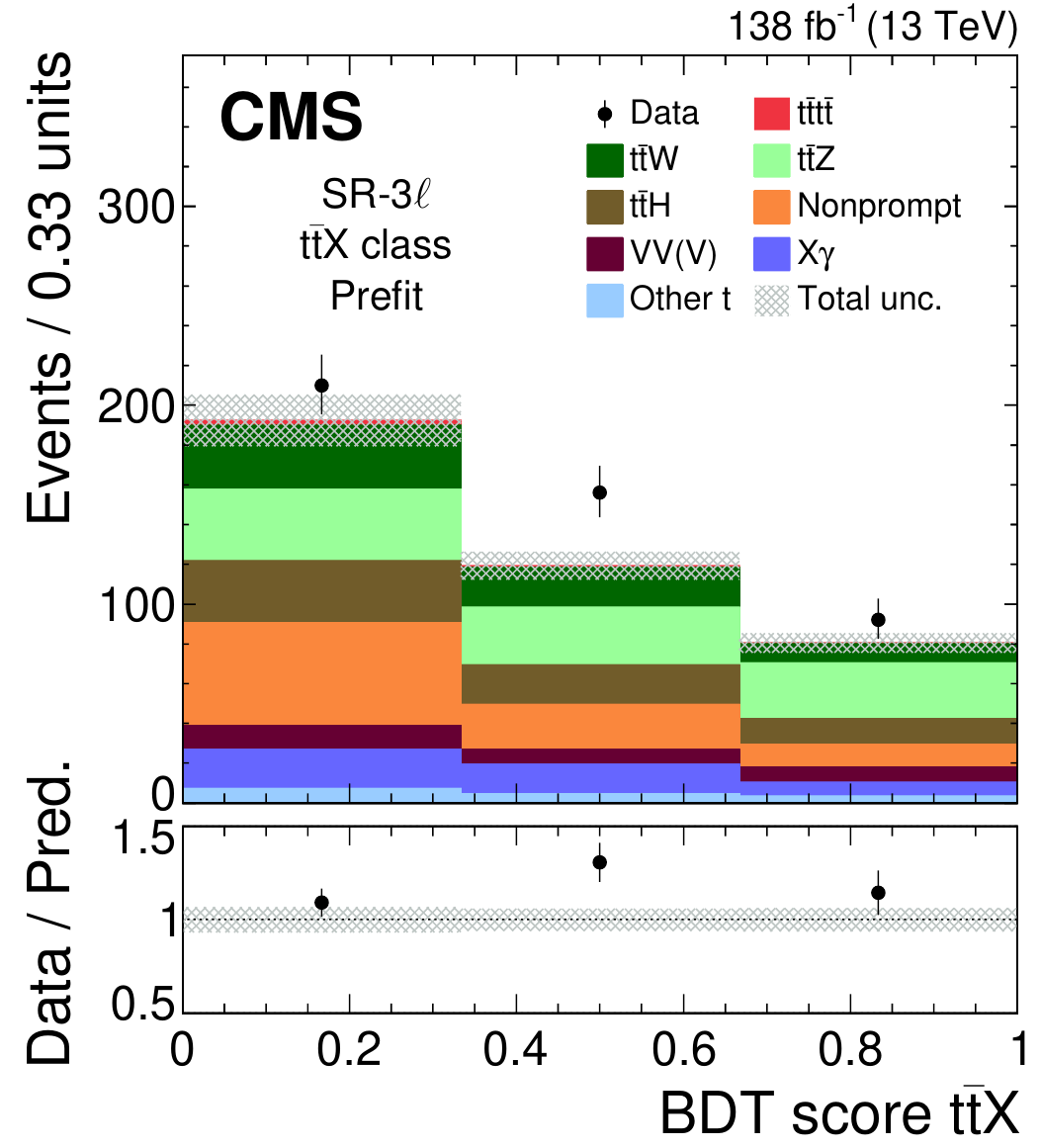}%
\hfill%
\includegraphics[width=0.3\textwidth]{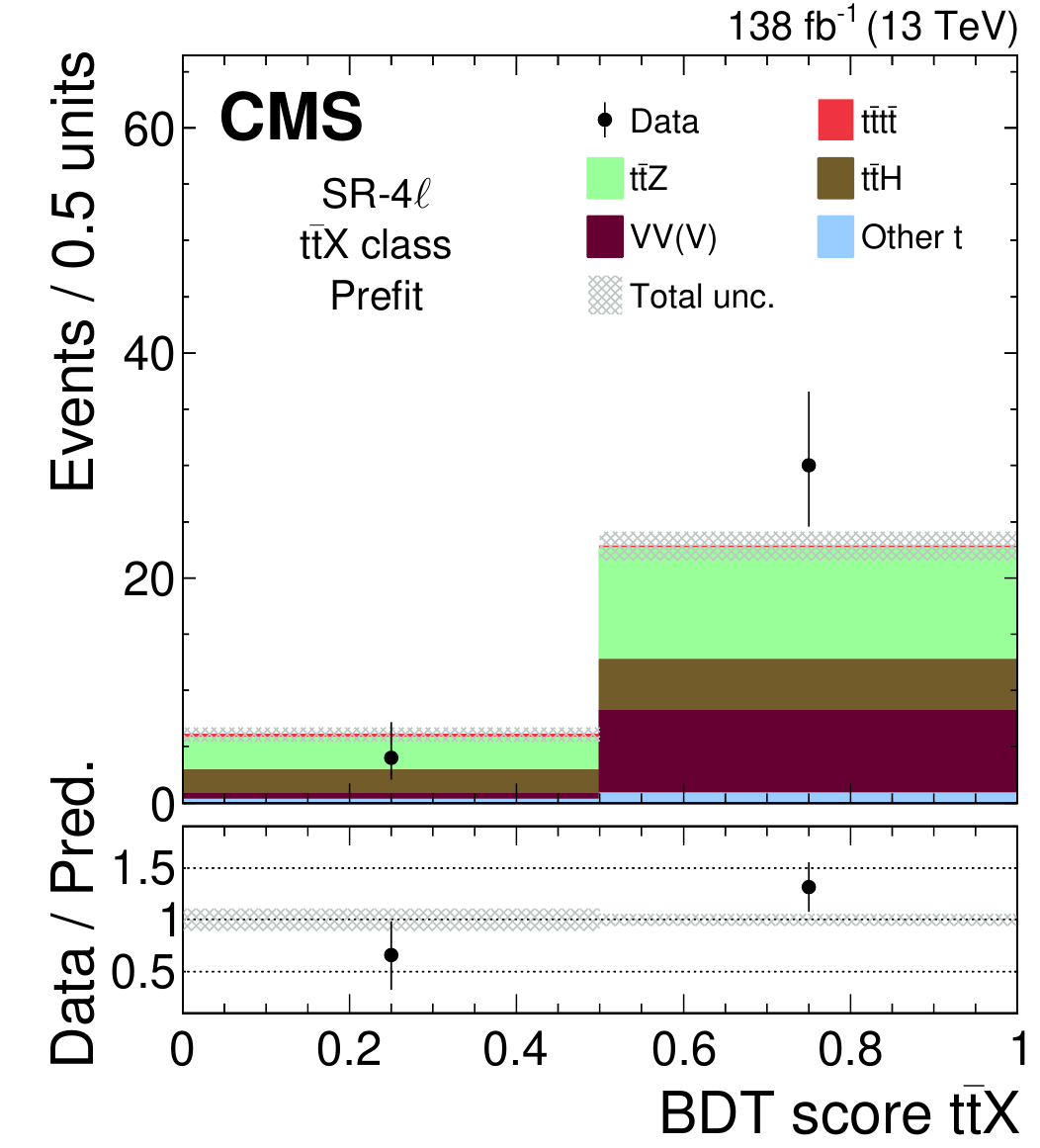}%
\caption{%
    Comparison of the number of observed (points) and predicted (colored histograms) events in the BDT score \ttX in the \ttX classes of \SRtwo in the \mumu category (left), of \SRthree (middle), and of \SRfour (right).
    The vertical bars on the points represent the statistical uncertainties in the data, and the hatched bands the total uncertainty in the predictions.
    The signal and background yields are shown before the fit to the data (``prefit'').
}
\label{fig:appendix1}
\end{figure*}

\begin{figure*}[!h]
\centering
\includegraphics[width=0.3\textwidth]{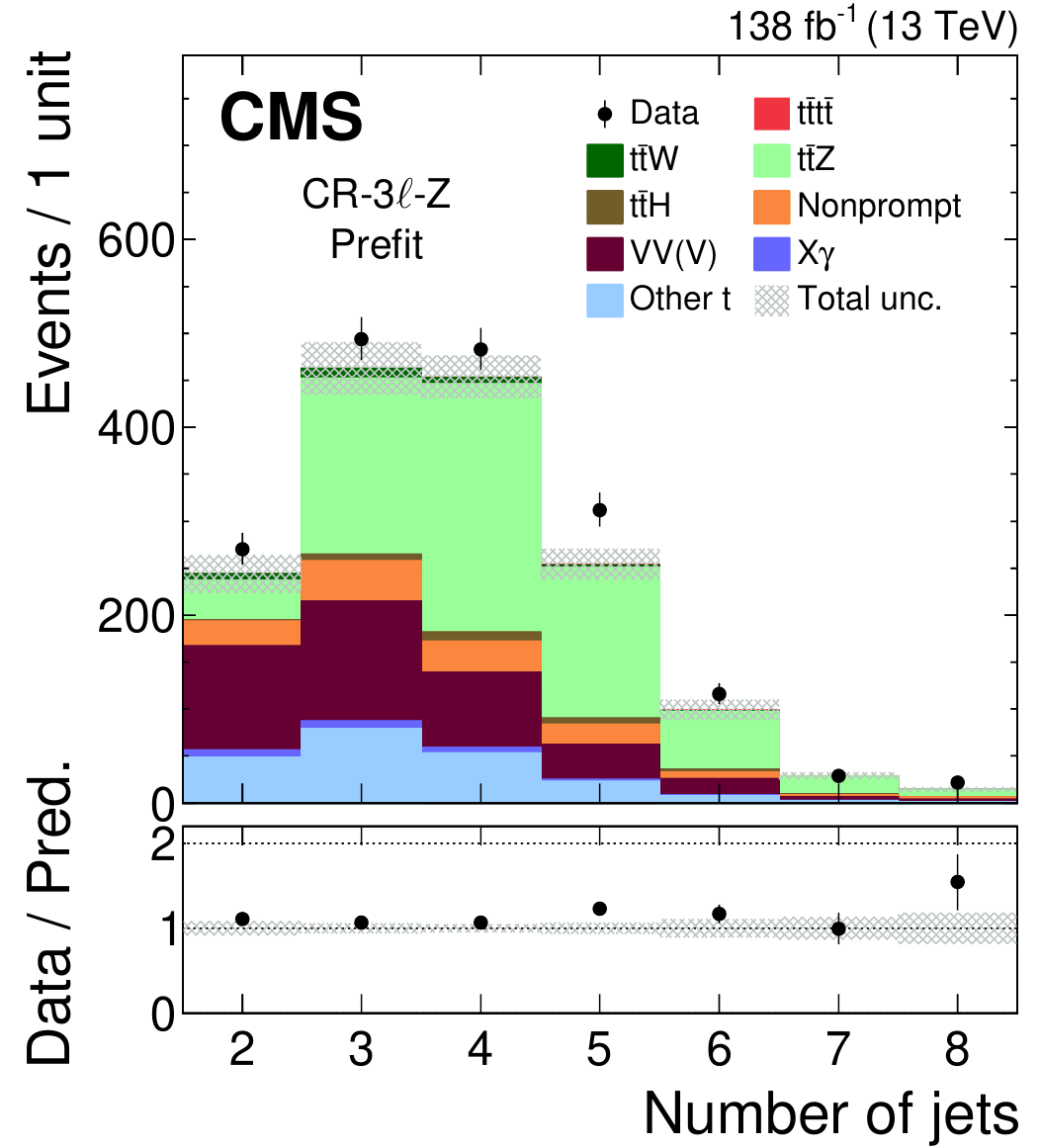}%
\hfill%
\includegraphics[width=0.3\textwidth]{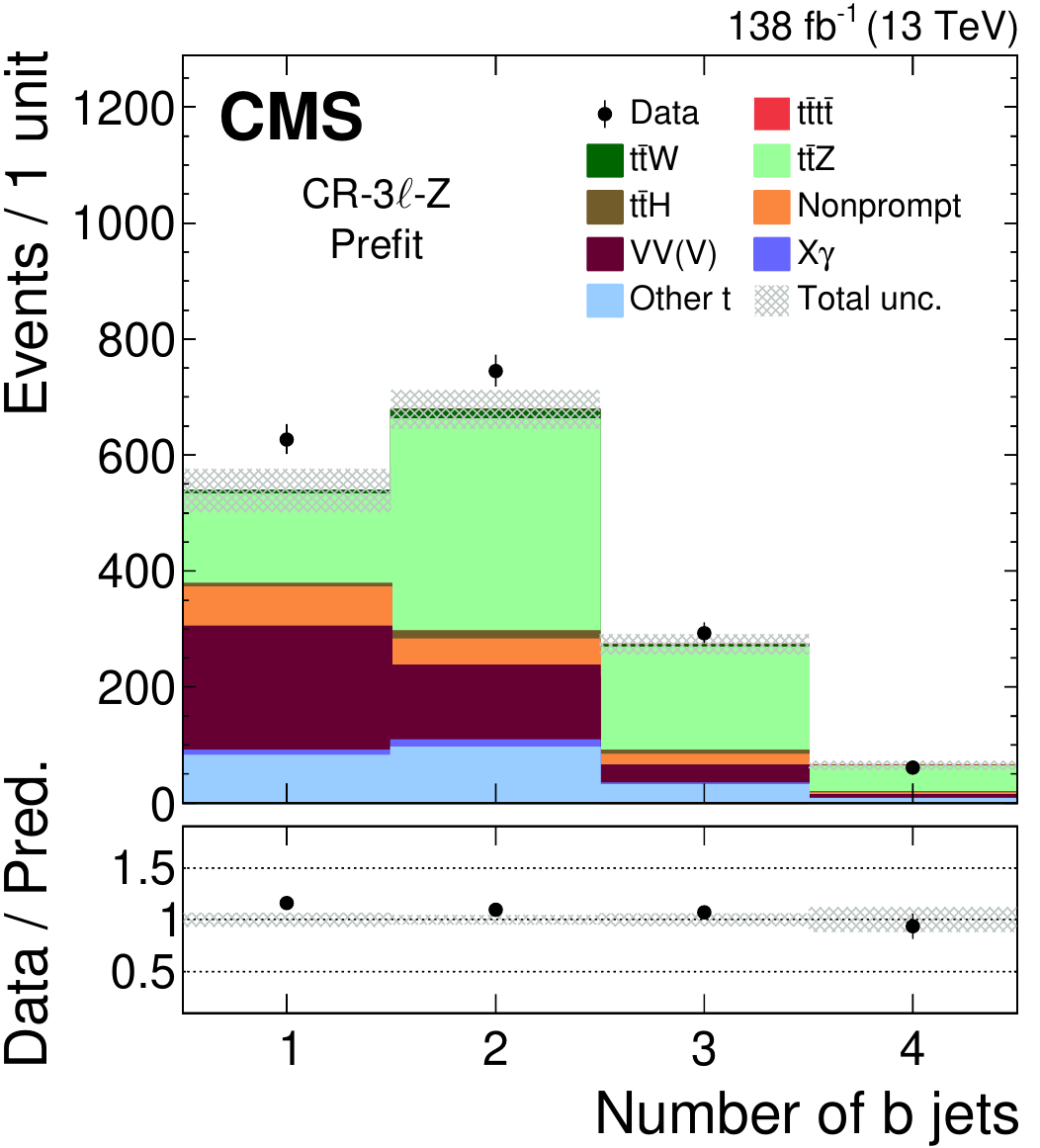}%
\hfill%
\includegraphics[width=0.3\textwidth]{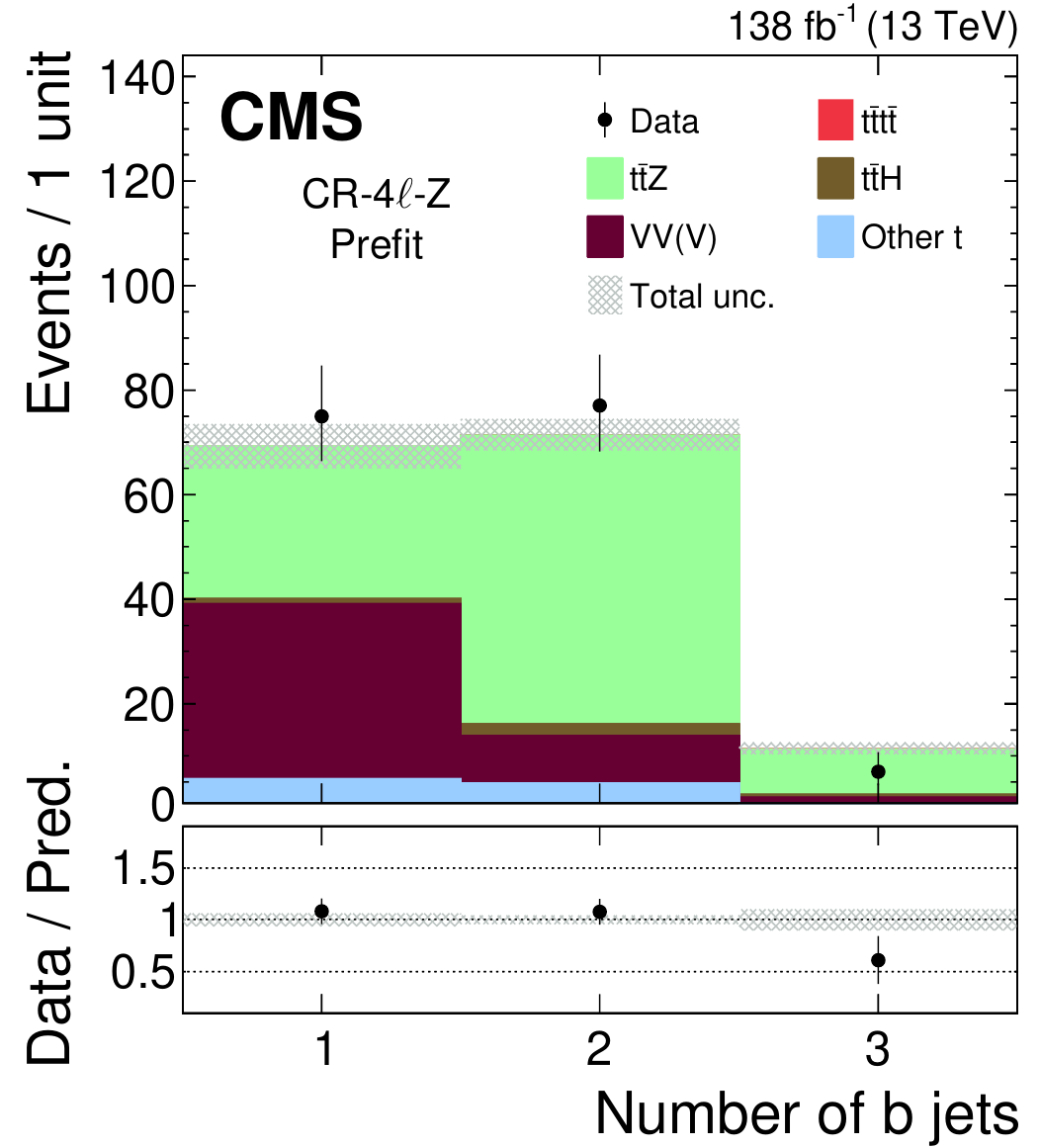}%
\caption{%
    Comparison of the number of observed (points) and predicted (colored histograms) events in the number of jets distribution in \CRthreeZ (left), and in the number of \PQb jets distribution in \CRthreeZ (middle) and \CRfourZ (right).
    The vertical bars on the points represent the statistical uncertainties in the data, and the hatched bands the total uncertainty in the predictions.
    The signal and background yields are shown before the fit to the data (``prefit'').
}
\end{figure*}

\begin{figure*}[!p]
\centering
\includegraphics[width=0.3\textwidth]{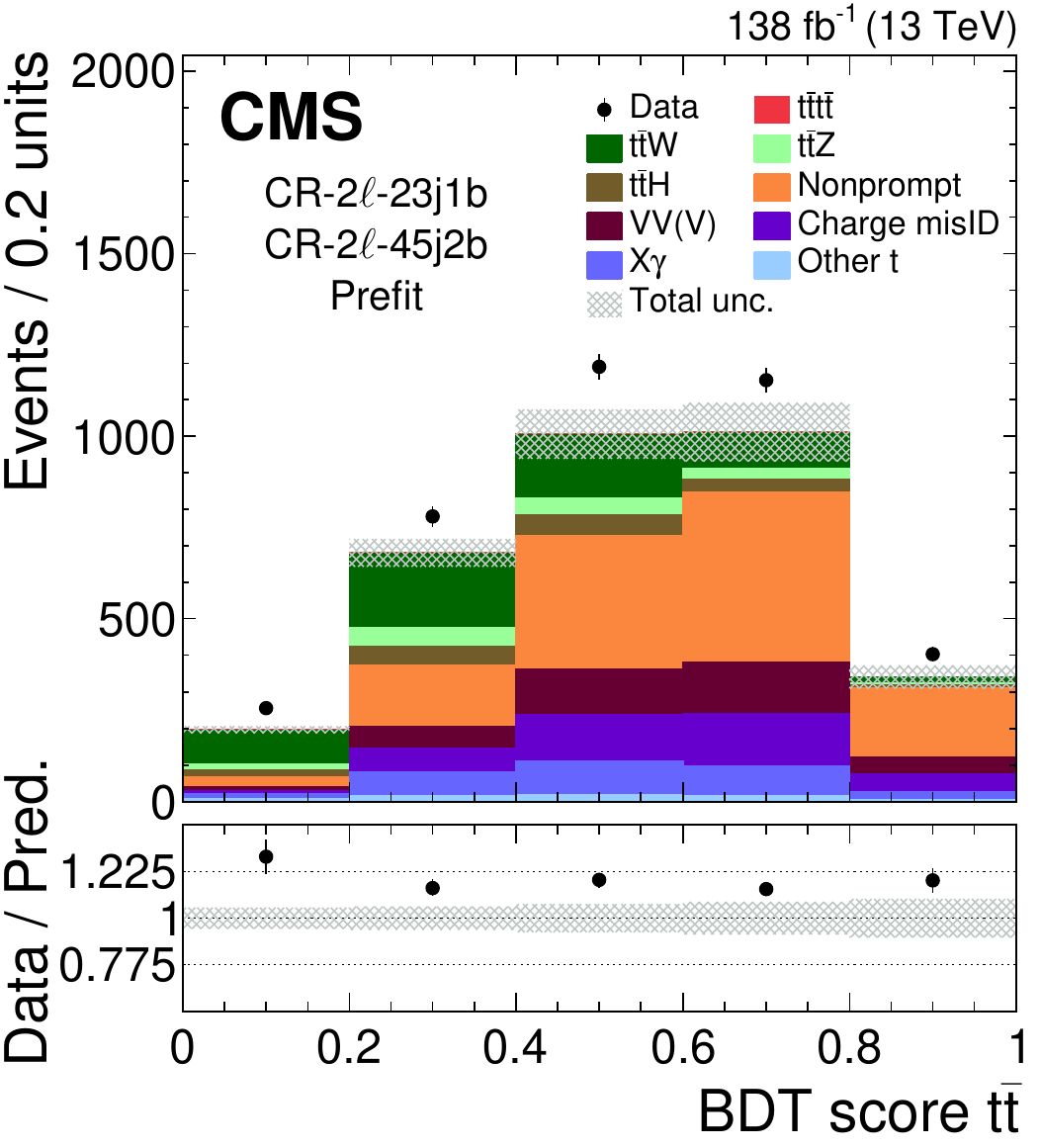}%
\hfill%
\includegraphics[width=0.3\textwidth]{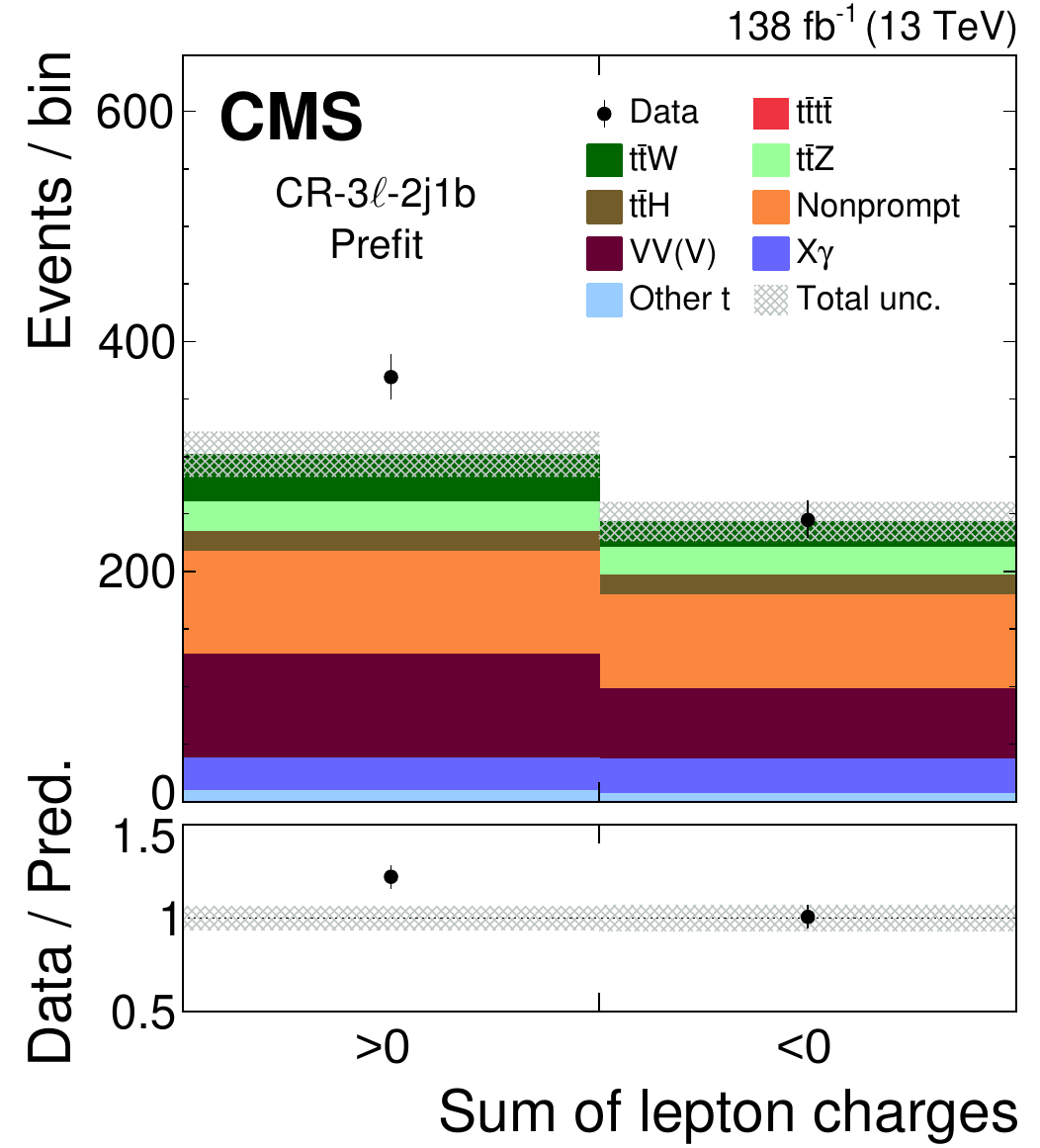}%
\hfill%
\includegraphics[width=0.3\textwidth]{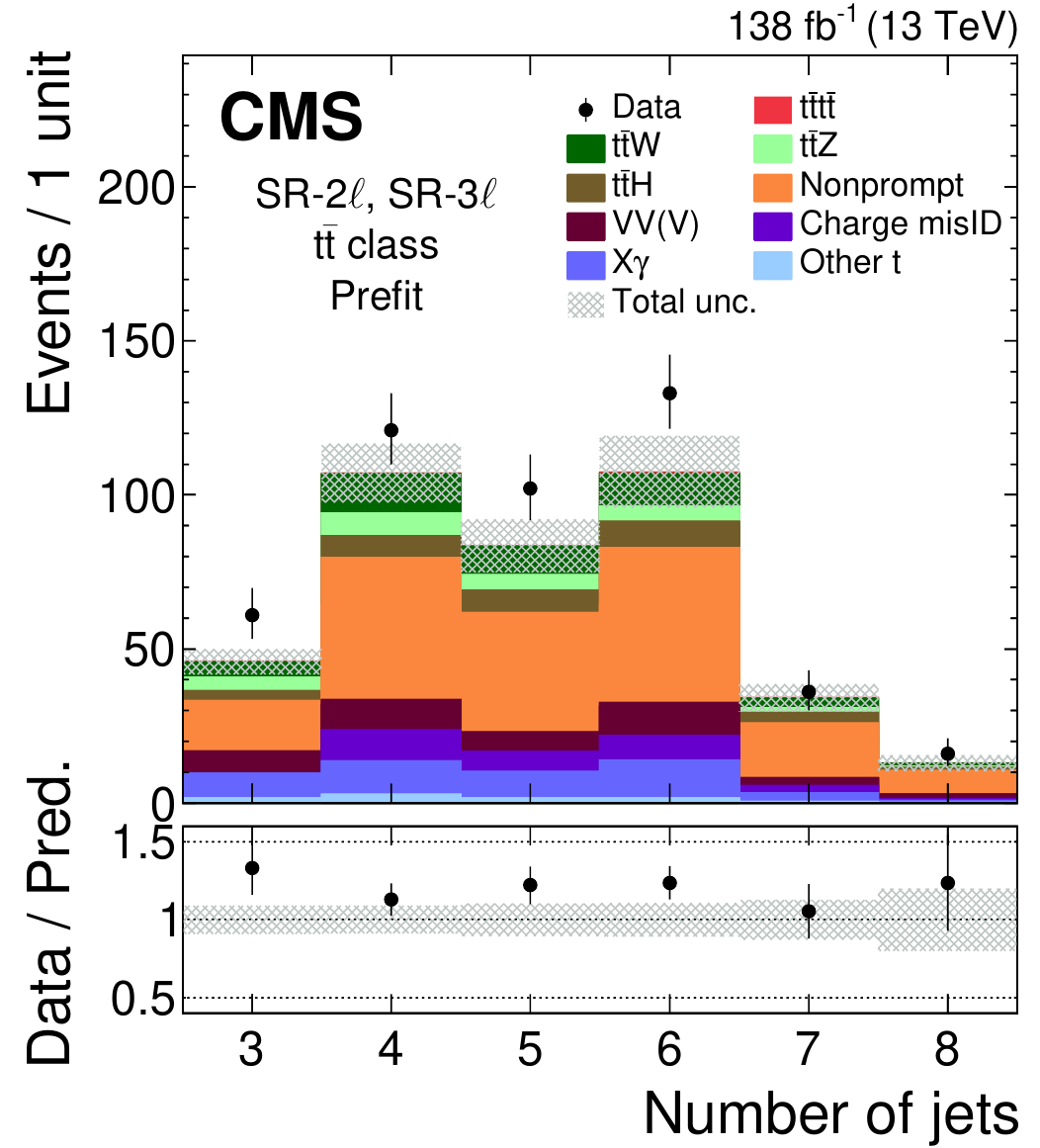}%
\caption{%
    Comparison of the number of observed (points) and predicted (colored histograms) events in the BDT score \ttbar in the combined \CRtwoLow and \CRtwoHigh (left), in the event yields with positive and negative sum of lepton charges in \CRthreeLow (middle), and in the number of jets distribution in the \ttbar class of the combined \SRtwo and \SRthree (right).
    The vertical bars on the points represent the statistical uncertainties in the data, and the hatched bands the total uncertainty in the predictions.
    The signal and background yields are shown before the fit to the data (``prefit'').
}
\end{figure*}

\begin{figure*}[!p]
\centering
\includegraphics[width=0.3\textwidth]{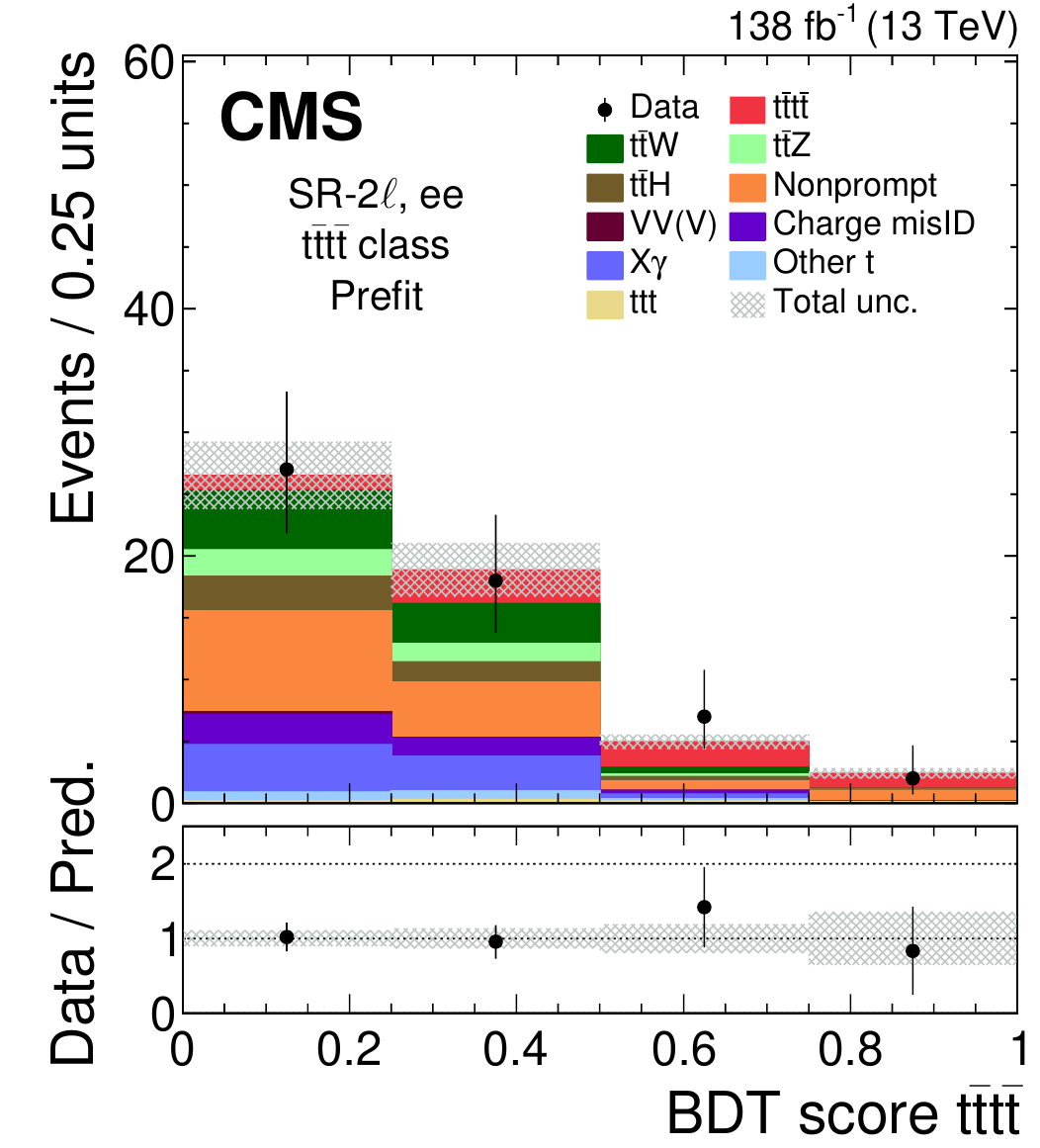}%
\hfill%
\includegraphics[width=0.3\textwidth]{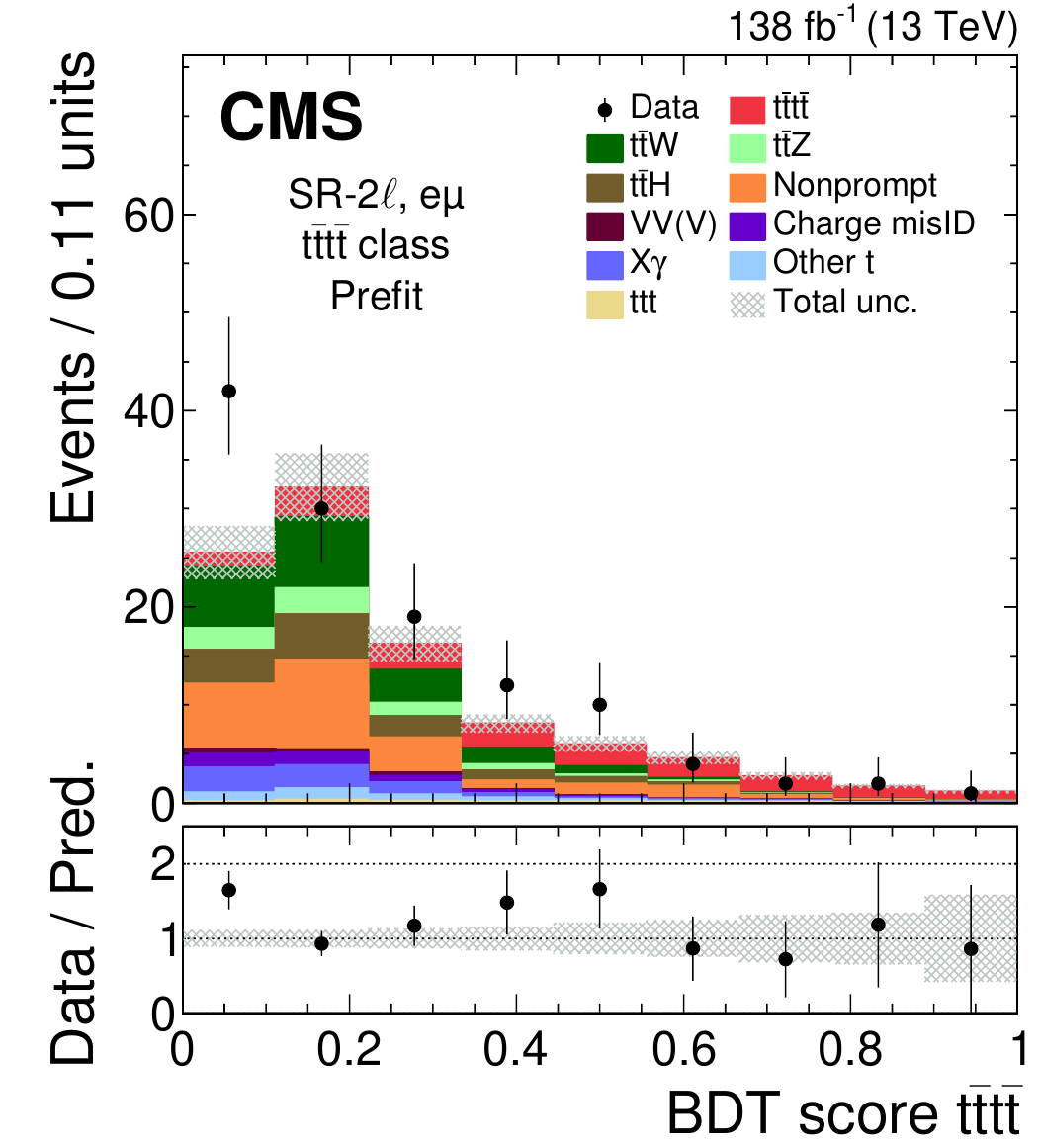}%
\hfill%
\includegraphics[width=0.3\textwidth]{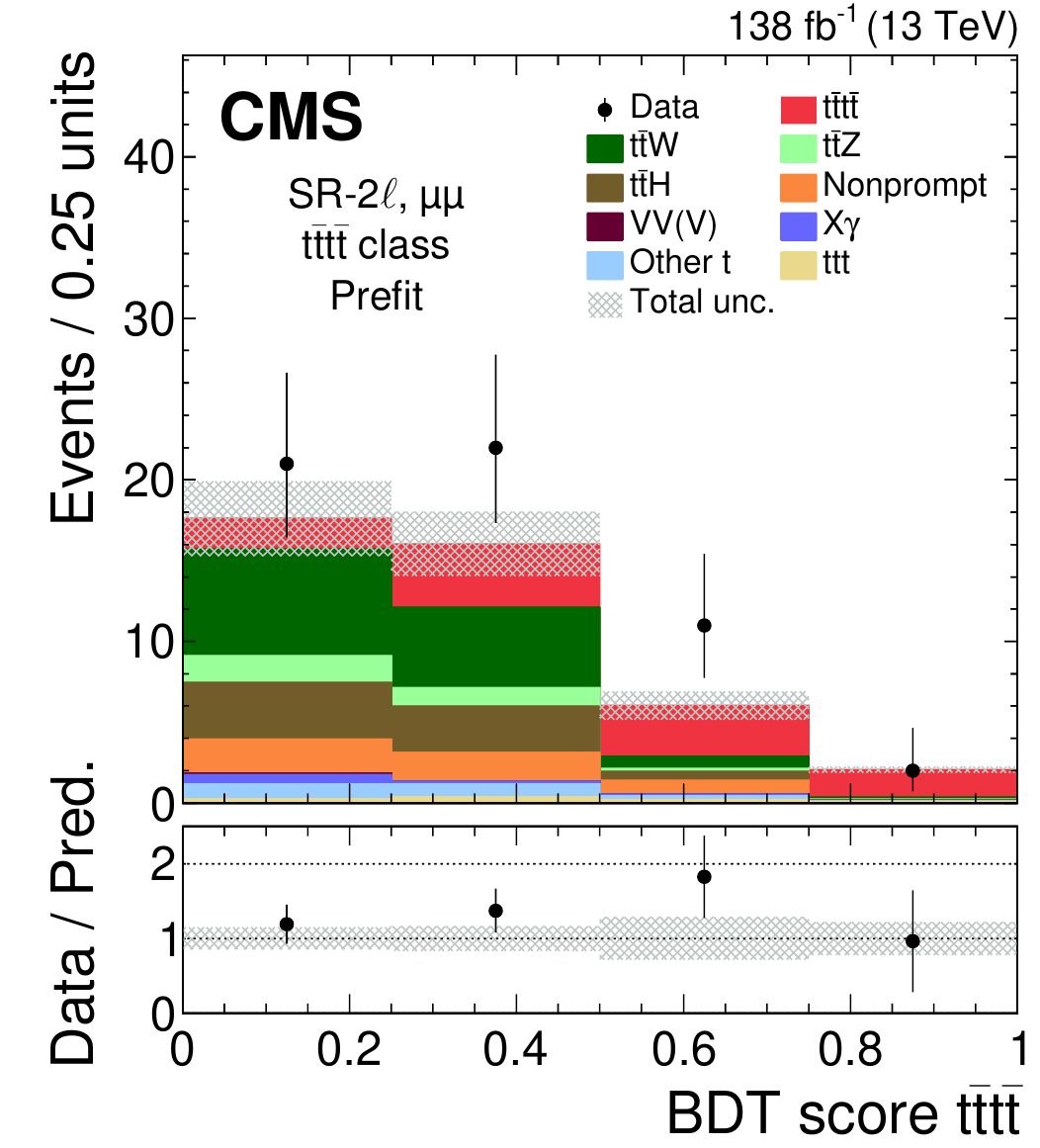} \\
\includegraphics[width=0.3\textwidth]{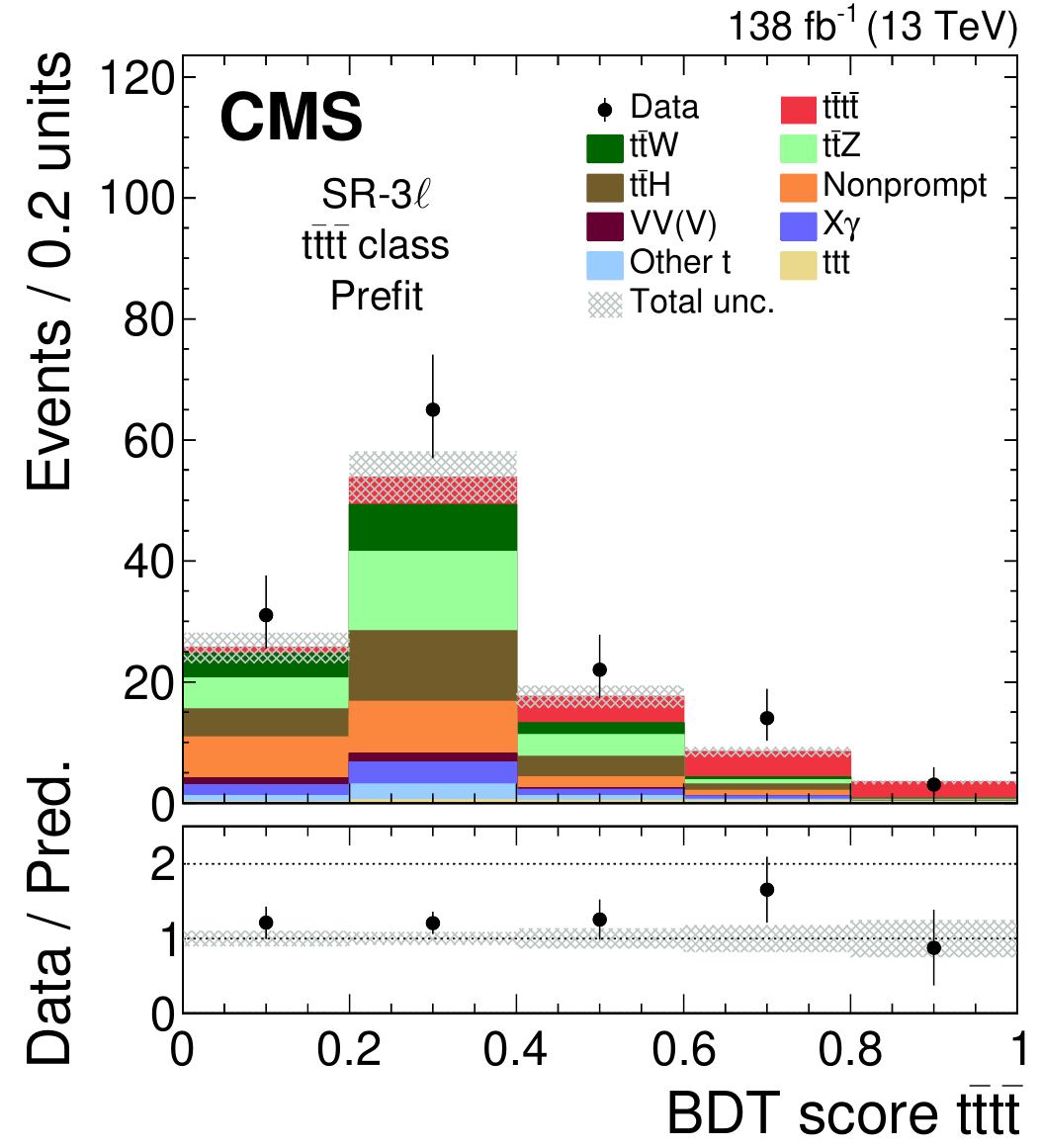}%
\hspace*{0.02\textwidth}%
\includegraphics[width=0.3\textwidth]{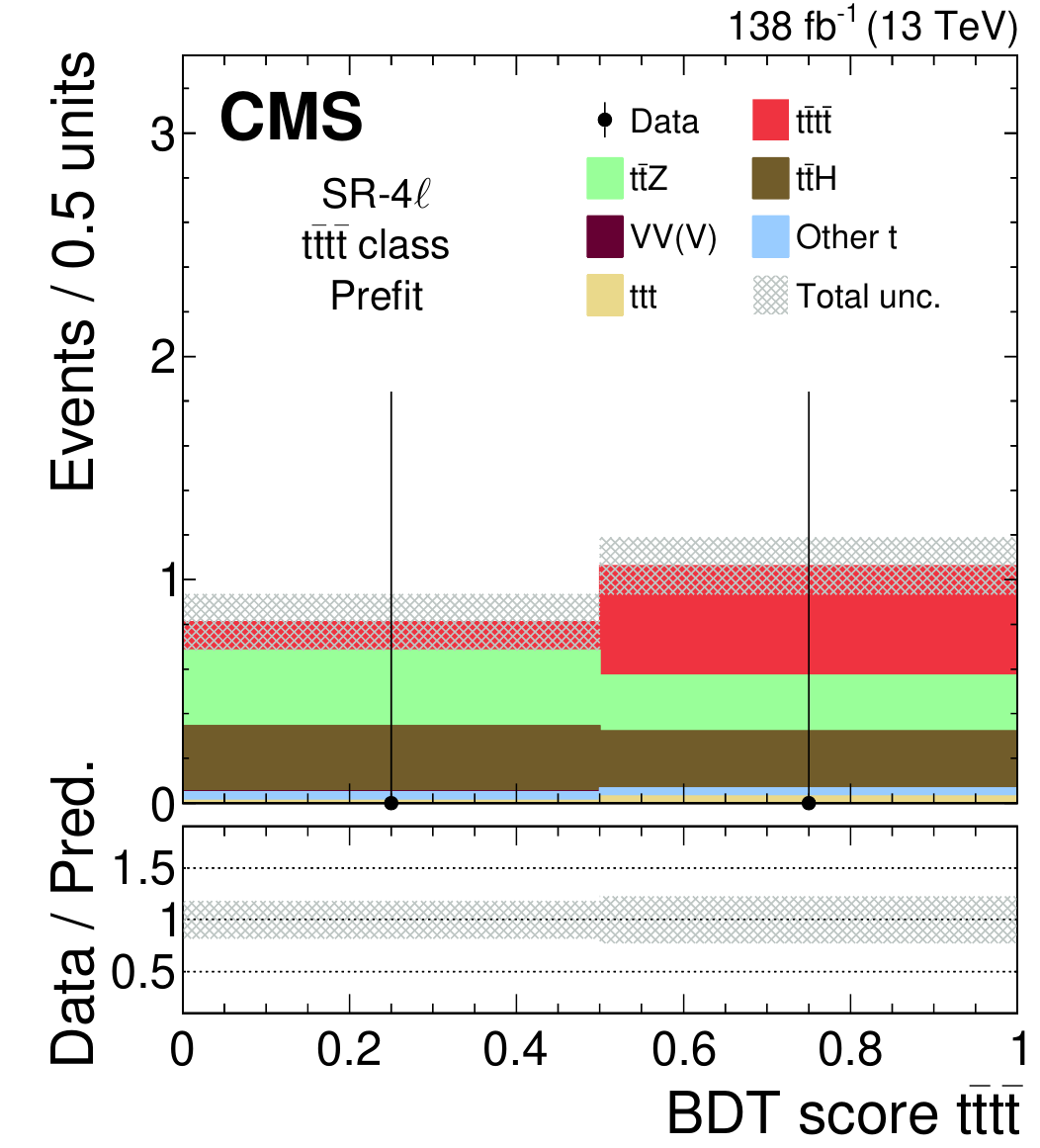}%
\caption{%
    Comparison of the number of observed (points) and predicted (colored histograms) events in the BDT score \tttt in the \tttt classes of \SRtwo, shown for the $\Pe\Pe$ (upper left), $\Pe\PGm$ (upper middle), and \mumu (upper right) categories, of \SRthree (lower left) and of \SRfour (lower right).
    The vertical bars on the points represent the statistical uncertainties in the data, and the hatched bands the total uncertainty in the predictions.
    The signal and background yields are shown before the fit to the data (``prefit'').
}
\label{fig:appendix4}
\end{figure*}
\cleardoublepage \section{The CMS Collaboration \label{app:collab}}\begin{sloppypar}\hyphenpenalty=5000\widowpenalty=500\clubpenalty=5000
\cmsinstitute{Yerevan Physics Institute, Yerevan, Armenia}
{\tolerance=6000
A.~Hayrapetyan, A.~Tumasyan\cmsAuthorMark{1}\cmsorcid{0009-0000-0684-6742}
\par}
\cmsinstitute{Institut f\"{u}r Hochenergiephysik, Vienna, Austria}
{\tolerance=6000
W.~Adam\cmsorcid{0000-0001-9099-4341}, J.W.~Andrejkovic, T.~Bergauer\cmsorcid{0000-0002-5786-0293}, S.~Chatterjee\cmsorcid{0000-0003-2660-0349}, K.~Damanakis\cmsorcid{0000-0001-5389-2872}, M.~Dragicevic\cmsorcid{0000-0003-1967-6783}, A.~Escalante~Del~Valle\cmsorcid{0000-0002-9702-6359}, P.S.~Hussain\cmsorcid{0000-0002-4825-5278}, M.~Jeitler\cmsAuthorMark{2}\cmsorcid{0000-0002-5141-9560}, N.~Krammer\cmsorcid{0000-0002-0548-0985}, D.~Liko\cmsorcid{0000-0002-3380-473X}, I.~Mikulec\cmsorcid{0000-0003-0385-2746}, J.~Schieck\cmsAuthorMark{2}\cmsorcid{0000-0002-1058-8093}, R.~Sch\"{o}fbeck\cmsorcid{0000-0002-2332-8784}, D.~Schwarz\cmsorcid{0000-0002-3821-7331}, M.~Sonawane\cmsorcid{0000-0003-0510-7010}, S.~Templ\cmsorcid{0000-0003-3137-5692}, W.~Waltenberger\cmsorcid{0000-0002-6215-7228}, C.-E.~Wulz\cmsAuthorMark{2}\cmsorcid{0000-0001-9226-5812}
% ...
\par}
\vskip\cmsinstskip
\dag:~Deceased\\
$^{1}$Also at Yerevan State University, Yerevan, Armenia\\
$^{2}$Also at TU Wien, Vienna, Austria\\
% ...
\end{sloppypar}
\end{document}